\documentclass[twocolumn]{emulateapj}
\usepackage{amsmath}
\usepackage{wasysym}
\usepackage{pstricks}

\slugcomment{Accepted for pblication in APJ 12/07/2013}

\shorttitle{ Modeling Statistical properties of ARs}
\shortauthors{Malapaka et al.}

\begin{document}


\title{ Modeling statistical properties of Solar Active Regions through DNS of 3D-MHD Turbulence}


\author{Shiva Kumar. Malapaka\altaffilmark{1,2} and Wolf-Christian M\"uller\altaffilmark{1,3}}
\affil{1: Max-Planck Institute for Plasmaphysics, Boltzmannstrasse 2, D-85748, Garching bei Muenchen, Germany. \\
 2: Department of Applied Mathematics, School of Mathematics, University of Leeds, Leeds, LS2 9JT, UK.\\
 3. Zentrum f\"ur Astronomie und Astrophysik, TU-Berlin, ER 3-2, Hardenbergstr. 36, 10623 Berlin.}




\begin{abstract}
Statistical properties of the Sun's photospheric turbulent magnetic
field, especially those of the Active Regions (ARs), have been studied
using the line-of-sight data from magnetograms taken by SOHO and
several other instruments (see e.g. \cite{abr02,abr03,abr10}). This
includes structure functions and their exponents, flatness curves and
correlation functions. In these works, the dependence of structure function
exponents ($\zeta_p$) of the order of the structure functions
($\it{p}$) was modeled using a non-intermittent K41 model. It is now
well known that the ARs are highly turbulent and are associated with
strong intermittent events. In this paper we compare some of the
observations from \cite{abr03} with the log-Poisson model
\citep{bis03} used for modeling intermittent MHD turbulent
flows. Next, we analyze the structure function data obtained from the
direct numerical simulations (DNS) of homogeneous,
incompressible 3D-MHD turbulence in three cases: sustained by forcing,
freely decaying and a flow initially driven and later
allowed to decay (case 3). 
The respective DNS replicate the properties seen in the plots of $\zeta_p$
against $\it{p}$ of ARs. We also reproduce the trends and
changes observed in intermittency in flatness [\cite{abr10}] and
correlation functions [\cite{abr03}] of ARs.\\ It is suggested
from this analysis that an AR in the onset phase of a flare can be
treated as a forced 3D-MHD turbulent system in its simplest form and
that the flaring stage is representative of decaying 3D-MHD
turbulence. It is also inferred that significant changes in
intermittency from the initial onset phase of a flare to its final
peak flaring phase, are related to the time taken
by the system to reach the initial onset phase.
\end{abstract}





\section{Introduction}
Active regions (ARs) of the Sun and their evolution have an impact on
weather, communications and sometimes health here on Earth. Sunspots,
flaring regions of the photosphere and coronal mass ejections (CMEs)
from the corona are some of the active regions which consist of highly
turbulent magnetized plasma \citep{arc12}. These are monitored using
highly sensitive instruments both from space and earth based solar
observations. Magnetograms obtained from these instruments provide
valuable knowledge of the physics and evolution of these regions
\citep{sti02}. Statistical properties of these turbulent ARs such as
the longitudinal structure functions, flatness and correlation lengths
have been studied using these magnetograms \citep{abr02, abr03, abr10}. 
The ARs analyzed in these works include the regions just
before a flare and during the flare (e.g NOAA AR 0039 in
\cite{abr03}). It was stated that the structure function exponent
curves for an AR before and during the flare a) differ, b) show
deviation from K41 and c) that this deviation is more significant
during the flare than before it \citep{abr03}. Some of the important
inferences obtained from this statistical analysis include the
understanding of the importance of small scales in the flaring process
\citep{abr02}, the probable physical mechanism involved in the flaring
process \citep{abr03} and possible multi-fractal nature of the
structures in the ARs \citep{abr10}.\\ It is now well-known that due
to the highly turbulent nature of these ARs, there exists
intermittency in these structures (see e.g. \cite{abr05, chi11}). Thus
it is prudent to involve intermittent models such as the log-Poisson
model, so as to have a better understanding of the interactions
between different scales in the flaring process. Also, there have been
no previous direct numerical simulations (DNS) of MHD turbulence that
capture the above-mentioned statistical behavior of the ARs. Hence,
this work focuses on comparing previously published data (from
\cite{abr03}) with a log-Poisson model and also on replicating the
observed trends using the simplest possible DNS of homogeneous,
incompressible 3D-MHD turbulence. Correlation functions and
flatness curves from the observations are also compared with the
simulation data. The principal aim of this paper is to show that it is
possible to replicate the observed statistical trends of ARs using
simple numerical simulations. Since the parameters of the simulations
can be controlled, further such attempts might in future result in an
exact match with the observations and hence might give us better
insight into the physical processes that cause solar phenomena such as
flares and CMEs. \\ This paper is divided into 5 sections. Section 2
describes the MHD equations, numerical simulation setup and the
statistical methods used to analyze the simulation data. Section 3.1
deals with the structure function exponent curves, with its two
subsections dealing with observational data and simulation data
respectively. Section 3.2 is dedicated to the discussion of flatness
curves and Section 3.3 for correlation functions. Discussion of the
statistical properties from both simulations and observations can be
seen in Section 4. The paper ends with Section 5 which lists the
conclusions reached in this work.
\section{Model equations, Numerical Setup and Statistical Methods} 
The dimensionless incompressible MHD equations giving a concise single-fluid description of a plasma are
\begin{subequations}
\label{solcon}
 \begin{equation}
 \partial_{t}{\boldsymbol{\omega}} =
 {\boldsymbol {\nabla}}\times({\boldsymbol{ v}}\times{\boldsymbol{\omega}} - {\boldsymbol{ b}}\times{\boldsymbol{j}})+
 {\mu}_{{n}}(-1)^{n/2-1}\nabla^{{n}}
 \boldsymbol{\omega}+{\boldsymbol{ {F_{v}}}}+{\lambda}\Delta^{-{1}}\boldsymbol{\omega}\,,
 \end{equation}
 \begin{equation}
\partial_{t}{\boldsymbol{b}} = {\boldsymbol {\nabla}}\times({\boldsymbol{v}} \times {\boldsymbol{b}}) +
 {\eta}_{{n}}(-1)^{n/2-1}\nabla^{{n}}
 {\boldsymbol{b}}+{\boldsymbol{{F_{b}}}}+{\lambda}\Delta^{-{1}}{\boldsymbol{b}}\,
\end{equation}
\begin{equation}
 {\boldsymbol{ {\nabla\cdot v}}} = {\boldsymbol{ {\nabla\cdot b}}}= {0},
\end{equation}
\end{subequations}
 with $\boldsymbol{\omega} = \boldsymbol{\nabla} \times
 \boldsymbol{v}$ the vorticity, $\boldsymbol{v}$ the velocity,
 $\boldsymbol{b}$ the magnetic field and $\boldsymbol{j}$ the current
 density. Relativistic effects are neglected. By defining the characteristic velocity 
 of the system to be the characteristic Alfv\'en speed and by giving 
 the magnetic field in Alfv\'en-speed units, the influence of 
 the specific numerical value of the characteristic constant density of the system 
 has been eliminated, for simplicity. This enforces the `interaction parameter' in front of the Lorentz-force term to be unity. Other effects such as
 convection, radiation and rotation are also neglected. Direct
 numerical simulations are performed by solving the set of model
 equations using a standard pseudospectral method \citep{can88} in
 combination with leap-frog integration on a cubic box of linear size
 $2\pi$ that is discretized with $1024$ collocation points in each
 spatial dimension. Spherical mode truncation is used for alleviating
 aliasing errors.  By solving the equations in Fourier space, the
 solenoidality of $\boldsymbol{ v}$ and $\boldsymbol{ b}$ is
 maintained algebraically.  Two main configurations, a driven system
 and a decaying one, are discussed here along with a third case in
 which the system is initially driven and later allowed to decay. In
 the driven case, the forcing terms ${\boldsymbol {F_v}}$ and
 ${\boldsymbol {F_b}}$ are delta-correlated random processes creating
 a small-scale background of fluctuations with an input of adjustable
 amount of magnetic and/or kinetic helicity. Additionally, a large
 scale energy sink\footnote{Since the system is forced at every
 time step, the peaks of the spectra moves to the lowest $k$ in a
 relatively short time. When they reach the largest wave-numbers, there can be
 a back reaction which pollutes the 
 spectral region where the scaling measurements of this study are carried out. 
 The energy sink used here is to avoid such consequences of finite
 spectral extent on the inverse cascade. The
 inverse Laplacian which in $k$-space transforms into $-k ^ {-2}$ and thus dominates dynamics at
 the lowest $k$. It is neither used in
 the decaying case nor in case 3 (once the system starts
 decaying). Further it is not be confused with an Ekman friction term
 used in 2D- hydrodynamic turbulence simulations (see \cite{bof02} and
 references therein), to inhibit the development of small scales;
 although $\lambda\Delta^{-1}$ closely resembles such a term. 
 }
 $\lambda\Delta^{-1}$ with $\lambda=0.5$ is present for both
 fields. In the decaying case the forcing terms and $\lambda$ are set
 to zero. The initial conditions for both the setups represent an
 ensemble of fluctuations having maximum magnetic helicity w.r.t. the
 energy content, with random phases having a smooth Gaussian
 distribution centered in intermediate scales for decaying case and in
 small scales for the driven case respectively. To reduce finite-size
 effects, the simulations are run for $6.7$ (forced) and $9.2$
 (decaying) large-eddy turnover times of the system respectively, with
 the time unit defined from the system size and its total
 energy. Since the size of largest eddies is only constrained by the
 physical boundaries of the flow \citep{fri95}, this chosen time scale
 encompasses all the significant interactions that occur between (and
 among) the structures of various scales (both small and large
 scales). \\

 In case 3, the simulation is continued for at least 5 further eddy turnover times from the instance the forcing is withdrawn. We study three sub cases of case 3:
\begin{itemize}
 \item{Case 3a : Data from the forced case at $t= 6.7$ is taken and allowed to decay, with the simulations stopped at $t= 11.34$}
 \item{Case 3b: Data from the forced case at $t=0.33$ is chosen and allowed to decay, with the simulations stopped at $t=5.79$}
 \item{Case 3c: Data from the forced case at $t=0.14$ is chosen and allowed to decay, with the simulations stopped at $t=6.26$}
 \end{itemize}
The initial state of case 3a [from here on case 3ai] is the final state of the forced simulation, while the initial state of case 3b [called case 3bi] is a randomly chosen state. The initial state of case 3c [from here on case 3ci] happens to be the instance of time when the peak of the spectrum of magnetic helicity, while inverse cascading, has moved to the spatial wavenumber $k=70$ from $k=206$. This state is similar to the initial state of the decaying system, with the major difference between these two systems being the ratio of energies (magnetic to kinetic), which is unity in the decaying case and $5.4$ in case 3ci. The final states of these simulations are represented as case 3af, case 3bf and case 3cf respectively. \\
 The hyperdiffusivities ${\mu}_{{n}}$ and ${\eta}_{{n}}$ are dimensionless dissipation coefficients of order $n$ (where $\it n$ is always even), with $n=8$ in all the runs described here, to obtain sufficient scale separation. The magnetic hyperdiffusive Prandtl number ${Pr_m}_n={\mu_n}/{\eta_n}$ is set to unity. It is difficult to define an unambiguous Reynolds number owing to the use of hyperviscosity \citep[][and the references therein]{mal09}. Further details of these equations and the numerical setup can be found in \cite{mul12, mulmal12}. It need to be emphasized here that, although the simulation setup described above will be used to understand the properties of ARs of the Sun, our current numerical setup is not capable of handling the density variations (compressibility effects) that are usually observed in these ARs. \\
 The real space data is obtained using inverse Fourier transforms to be processed for the statistical analysis. 
\subsection{Statistical analysis of Structures}
The mathematical formulation for calculating the three statistical properties, namely structure function exponent, flatness curves and correlation functions,for use with both observational and simulation data, is described below.
\subsubsection{Structure Functions}
Longitudinal structure functions of the magnetic field $S^{\it{b}}_{\it{p}}(\ell)$ in a sphere of radius $\ell$ are defined; following \cite{pol95,hor97,bis03}, as :
\begin{subequations}
\begin{equation}
\delta\it{b_{\ell}} = [\bf{b}(\bf{r}+\boldsymbol{\ell}) - \bf{b}(\bf{r})]\cdot \boldsymbol{\ell}/\ell
\end{equation}
\begin{equation}
\it{{b}_{\ell}} = \left\langle \delta\it{b_{\ell}}^{2}\right\rangle^{1/2}.
\end{equation}
\begin{equation}
S^{\it{b}}_{\it{p}}(\ell) = \left\langle \delta\it{b_{\ell}}^{\it{p}}\right\rangle \sim  \ell^{\zeta_{\it{p}}},
\end{equation}
\end{subequations}
\begin{equation}
\zeta_{p} = (1-x)p/g+\it{C_{0}}(1-(1-x/\it{C_{0}})^{p/g});
\end{equation}
where $\bf{r}$ and $\delta\it{b_{\ell}}$ are the radius vector and the longitudinal fluctuations in the magnetic field over a sampled distance $\ell$, respectively. For simplicity, these fluctuations are assumed to only depend on $\ell$ (as represented by equation 2b). Here $\zeta_{\it{p}}$ is a $\it{p}$-dependent scaling exponent and $\bf{b}$ is the magnetic field, where $\it{p}$ represents the `order' and $\delta\it{b_{\ell}}^{\it{p}}$ the longitudinal fluctuations in the magnetic field at that order. Note that the above equation (2c) is only valid in the self-similar (scaling) range(s) of the spectra.\\
Two methods, namely local-slope analysis [LSA] \citep{per09, sah11} and extended self similarity analysis (ESS) \citep{ben93,bis00,bis03} will be explored in order to determine the value of $\zeta_{\it{p}}$. To model $\zeta_{\it{p}}$ equation (3) is used, where parameters $\it{x}$, $\it{g}$, related to the basic spatial scaling of the turbulent fields, $\sim\ell^{1/g}$, and $\it{C_{0}}$, 
the co-dimension of the most singular dissipative structures in the flow,  are determined on physical and phenomenological grounds. For 3D-MHD turbulence, $\it{x}=2/\it{g}$ where $\it{g}=3$ in Kolmogorov phenomenology and $\it{x}=2/\it{g}$ where $\it{g}=4$ in Iroshnikov-Kraichnan phenomenology. $\it{C_{0}} = 1$ represents two-dimensional structures (sheets), $\it{C_{0}} = 2$ represents one dimensional structures (filaments) \citep{fri95,bis03}. For supersonic driven MHD turbulence, a variant of the log-Poisson model, to understand the velocity structure functions, also exists \citep{bol02}. However, since the focus of this work is on understanding the magnetic field structures, this model is not relevant here in this paper. Thus, in the work described below, we only use model curves obtained using Kolmogorov phenomenology for  3D-MHD turbulent flow, as they appear to fit the data the best.
\subsubsection{ Flatness}
 Flatness $F(\it{\ell})$, is defined as:
\begin{equation}
F(\it{\ell}) = \frac{S_{6}(\it{\ell})}{S_{2}^{3}(\it{\ell})} \sim \ell^{- \kappa}
\end{equation}
where $S_{6}$ and $S_{2}$ are the sixth order and second order moments respectively, following \cite{abr10}. The flatness function has a power law dependence on the scale $\ell$, with the flatness function exponent (or simply flatness exponent) of $\kappa$ characterizing the degree of intermittency \citep{abr10}. For non-intermittent structures determined by a Gaussian, F($\it{\ell}$) has a value of `3' and any deviation from this value is a  measure of intermittent features in the field.
\subsubsection{Correlation Functions}
The correlation function $\rho(\it{\ell})$ for the magnetic field is defined as:\vspace{1.5mm}
\begin{equation}\label{5.8}
\rho(\it{\ell})=\frac{\int^{X}_{-X}{\it{b_{x}}(\it{x})\it{b_{x}}(\it{x+\ell})\it{dx}}}{\int^{X}_{-X}{\it{b_{x}}^{2}(\it{x})\it{dx}}}
\end{equation}
where `X' is taken to be much greater than any characteristic length scale associated with the fluctuations in $\it{b_{x}}$. Here it represents the extent of the simulation box. The terms $\it{b_{x}}$ and $\it{b_{x}}(\it{x+\ell})$ are fluctuations in the field at two points $x$ and $x+\ell$.

\section{Results}
\subsection{Structure Function Exponent Plots}
\subsubsection{Remodeling of plots of $\zeta_{p}$ against  $\it{p}$ for ARs}
Values of both $\zeta_{p}$ and $\it{p}$ of the magnetic field, are extracted from the plots of fig.2 of \cite{abr03} which are tabulated in table. 2 (see appendix) and are plotted in fig.1b along with the original figure in fig.1a. As can be observed from fig. 1a, it is clear that data does not comply with the non-intermittent model curve used for comparison. However, from the remodeled data of fig.1b, it is seen that the $\zeta_p$ values before the flare (blue data points) fit well with the model curve indicative of the multi-fractal nature of the structures in this phase, while the $\zeta_p$ values during the onset of the flare (red data points) fit well with the intermittent model curve [with the expected deviations from the model curve observed for higher-order statistics (see sub-section below for more details)], indicative of sheet-like structures. Thus it appears that intermittency in the structures of ARs increases during the flaring stage, along with a change in the dimensionality of the structures.
\subsubsection{ The relation between $\zeta_{p}$ and $\it{p}$ from simulations}
The structure functions, up to order 8, of the magnetic field from all three cases (forced, decaying and case 3af) have been calculated using the equations (2a) to (2c), described above. LSA and ESS analysis of this data is performed to obtain $\zeta_p$ values for all studied cases and these values are tabulated in tables 3, 4 and 5 along with their respective errors (see Appendix for plots, the details of the  calculation procedure and the tables).\\
 These tabulated values are plotted in figs. 1c and 1d, where the former corresponds to $\zeta_p$ values obtained from LSA and the latter from ESS. Model curves corresponding to different values of $C_0$ along with a non-intermittent model curve are also plotted in these graphs for comparison (see figure caption for details). As can be observed from the data points in figures 1c and 1d, the information obtained from both ESS and LSA is extremely similar. However, LSA gives better error estimates than ESS, providing additional confidence on the data, as was observed by \cite{sah11}. Although the $\zeta_p$ values up to order 4 for all the cases are very close to each other and do not yield much information on the nature of the structures, the higher order structure function exponents do give useful information. Since we start our simulations with random phases and Gaussian fluctuations which are non-intermittent, the $\zeta_p$ values of the initial state of the simulations (not shown here) would have coincided with the black dotted line of figs. 1c and 1d . As the simulations progress, the structures quickly (in less than one large eddy turnover time) start to become intermittent (see data points for case 3ci [orange $\times$] and case 3bi [magenta $\square$] in figs. 1c and 1d) and are filament-like at these instances in the forced system. By the time the simulation ends, in the forced case, the structures indicate fractal dimensions (1.5 to be specific), as can be inferred from the forced / case 3ai data [blue $\circ$]. The observational data of the ARs before the onset of the flare NOAA AR 039 (plotted in figs.1a and 1b) shows similar behavior to that seen in the forced / case 3ai system. \\
 The $\zeta_p$ values for cases 3bf and 3cf (green $\diamond$  and maroon $\vartriangle$ respectively) are just above the data points for the decaying case (red $\hexagon$), and those for case 3af (black $*$) are just below them, at the higher orders, in figs. 1c and 1d. All these data points lie close to the intermittent model curve with $\it{C_{0}} = 1$, representing sheet-like structures. These curves show similar behavior to the  observed $\zeta_p$ values of the ARs during the rising phase of the flare (see fig. 1a and 1b) from NOAA AR 039. \\
 An interesting thing to note from these plots is that the $\zeta_p$ values at higher orders ($p > 4$) follow the sequence of time at which these values have been calculated from case 3ci, which is the earliest instance of time, to case 3af, which is the farthest instance of time, with the rest of the cases lying in between these two sets of curves. It is also interesting to note that the deviation from the model curve that represents sheet-like structures also depends on the same factor. Case 3bf, in which the simulation was stopped earliest, shows less deviation from the model curve, with case 3cf, the decaying case and case 3af showing increasing deviations in that order.\\
 The transition of the structure function exponent curves from cases 3ai, 3bi and 3ci to cases 3af, 3bf and 3cf is precisely the behavior that is observed in the structure function exponent curves of the ARs before and during the flare. These simulations also capture the inherent fractal nature of the structures before the flare in the ARs. It is important to note that our simulations are performed for a highly idealized spatially homogeneous system with no characteristic internal length scale which prevents a direct comparison of the structure functions obtained numerically and the ones calculated from the observations of ARs without, for example, introducing an arbitrary scale transformation in $\ell$. However, we do not know of any other past work, that mimics the observed trends of the $\zeta_p$ values of ARs so closely hence lending the presented results from our simulations relevance.\\  
 Although the structure function exponent curve method helps in determining and understanding the nature of intermittency in the structures of a flow, it does not quantify this intermittency. Flatness curves can help us quantify the level of intermittency \citep{tsi01} and hence are discussed next. 
 \subsection{Flatness Curves}
 Flatness curves of the magnetic field, for all the cases discussed above are plotted in fig. 2c, using the equation (4) from the previous section. A selection of the flatness curves plotted for different ARs in \cite{abr10} are shown in fig. 2a and 2b for reference. In \cite{abr10} the intermittency spectra of ARs are divided into three types (a) type 1: highly intermittent ARs with pronounced, steep power laws extending almost a decade on the logarithmic x-axis (size of the structures), (b) type 2: the ARs where the power law region is not so pronounced and the flatness exponent is smaller than that of the type 1 ARs and (c) type 3: ARs which show no intermittency.\\
 In the simulations, the cases 3bi [solid dark khaki line] and 3ci [dotted maroon line], which are early states of the forced simulation, show a low level of intermittency . However, as time progresses, in the forced case, intermittency increases and by the time the forced simulation is stopped i.e. forced / case 3ai, the flatness curve (thick red line) shows a linear region which satisfies a power law of $\kappa = -0.15$ (the dotted black line over the red line). The decaying case curve (thick blue line) shows a predominantly broad spectral power range with a $\kappa = -1.56$ (the dotted black line over the blue line). case 3af (thick dark green line) shows the broadest spectral range with $\kappa = -1.7$ (dotted black line over the dark green line). A surprising trend is seen in the curves of case 3bf (thin maroon line) and case 3cf (thin orange line), with the curve of case 3cf falling below the curve of case 3bf, although the simulation for case 3cf was run for a longer time. The other important point to note is that the value of $\kappa$ varies significantly from -0.01 (case 3bi) to -1.7 (case 3af), with values for other cases residing between these two limits (for clarity only 3 such fits are shown). It can be deduced from the plots of the simulated cases that these flatness curves show similar trends to their counterparts for ARs. Case 3af has highest intermittency, and agrees with the profiles of highly intermittent ARs, while the forced case agrees with type 2 ARs. Since all of the simulated states are intermittent, we did not find type 3 ARs in our simulations, however case 3bi and case 3ci have very small values of $\kappa$, indicating moderately intermittent states.
\subsection{Correlation Functions}
The correlation functions are a convenient geometrical parameter for detecting a strong intermittent event, as stated in the discussion of the fig.4 of \cite{abr03}. Thus we plot correlation functions of the magnetic field for all the simulated cases to see if we can replicate the observed trend.
 The correlation functions of the studied cases do not show specific patterns / relations from one case to another, except that when the forcing is withdrawn and the system is allowed to decay the correlation function moves to the right of the correlation function of the initial state. This pattern is seen in fig. 3b for cases 3ci (thin khaki line), 3bi (dotted maroon line) and Forced /3ai (thick red line), to the right of which are the final states of these cases 3cf (thin orange line), 3bf (thin magenta line) and 3af (thick green line) respectively. The correlation function of the decaying case (dotted blue line) is seen just after case 3bf and below the forced case. Figure. 3a (the original figure 4 of \cite{abr03}) shows the correlation function plots of NOAA AR 9661, with the correlation function near the peak of the flare (thin black line) appearing to the right of the correlation function of the beginning of the flare (thick black line). Thus it appears to be prudent to consider the beginning of the flare as the end (of some form of) a driven flow, while the peak state of the flare could be considered as a decaying flow, with greater intermittency than the initial forced state.
 \section{Discussion}
The plots of the studied cases shown, along with the related plots of ARs, in the above section establish the idea that the statistical properties of ARs could be replicated in simulations. Further, it is interesting to study the influence of other turbulent parameters on the observed statistical behavior of the simulation data, to see if any useful inferences could be drawn to improve our understanding of the statistical behavior of the ARs.\\
First, we see the influence of initial conditions and how this changes in each of the studied simulations, by tabulating the values of magnetic energy $E_M$ and kinetic energy $E_V$ at the initial state (I /i) and final state (F/f) in table 1 for each of the cases. We then calculate the ratio ($\frac{F}{I}$) for each of these cases and tabulate these values as $E_{MR}$, $E_{VR}$, $E_{TR}$ and $H_{MR}$ to represent the change in ratio of $E_M$, $E_V$, $E_T$ (total energy =  $E_M$ + $E_V$) and magnetic helicity $H_M$, from their initial state value to final state value; respectively \footnote{These quantities are given a subscript so as to differentiate from their respective spectral quantities which usually have a super script.}. At each calculated value an arrow is also shown with an $\uparrow$ /$\downarrow$  indicating an increase / decrease of the quantity from its initial value. In the forced case, the values of $E_V$, $E_T$, $E_M$ and $H_M$ increase by  a factor of 3.5, 9, 13.7 and 33 respectively from their initial values. The final state of the forced case is the starting point for the case 3a (i.e. case 3ai), and by the time this simulation is stopped (case 3af), these values fall by a factor of 5, 0.55, 1.6 and 0.01 respectively, from the case 3ai values. If we look at the structure function plots and flatness curves, the change of intermittency from Forced / case 3ai to case 3af is more pronounced than in any other case. If we look at table 1 for the cases 3b and 3c; the depreciation of these values  between the initial states case 3bi and 3ci and their respective final states case 3bf and case 3cf is of the same order and the change in intermittency is also similar. In the decaying case, where we start with higher initial energies (both kinetic and magnetic), the depreciations are much larger for these values with $E_{VR}$, $E_{TR}$, $E_{MR}$ and $H_{MR}$, having values of 185, 61, 37 and 0.05 respectively. The intermittency in this case is however higher than that of the cases 3bf and 3cf but is smaller than 3af.\\
It is seen in \cite{mul12}, that magnetic helicity shows a power law of $k^{-3.3}$ in the forced case (at large scales) and $k^{-3.6}$ in the decaying case, while no change in the power law of magnetic energy is observed, it stays at $k^{-2.1}$ for both cases (i.e. at large scales of the forced case and the decaying case). When we now plot the magnetic helicity spectra of cases 3ai and 3af, we observe that the power law of this quantity changes from $-3.3$ in the initial to $-3.6$ in the final state (see fig. 4a), however the power law for magnetic energy does not vary from $-2.1$ (fig. 4b). Except for magnetic helicity, the spectrum of no other quantity discussed in relation (6) of \cite{mul12}, changes so significantly. Thus it is to be believed that the observed change in intermittency is associated with a change in the topology of the underlying flow from the forced case to the decaying case, as magnetic helicity is a topological property of the flow. No such comment could be made for cases 3b and 3c as the initial state is not quasi-stationary enough to look for any power law behavior in the spectral space, although their final state spectra do coincide with a $-3.6$ power law of the decaying case. The realization that topological changes have an important role in the evolution of statistical properties of ARs was already reached by \cite{sor04} where the analyzed parameters are partition function exponents and cancellation exponents of both current helicity ($H^{J} \sim k^{2} H^{M}$) and current ($\bf{j}$), with support from DNS. Since current helicity and magnetic helicity are related spectrally, if the spectra of current helicity before and during a flare are obtained from magnetograms, the above-mentioned change in power law could be verified in proxy. Further, as magnetic energy, kinetic energy, magnetic / current helicity and kinetic helicity are also spectrally related (see \cite{mul12, mulmal12}), if observations are able to obtain the data of one or more of these parameters aside from current helicity, it can help in understanding the interplay of these quantities in the ARs at different phases, thus improving our knowledge of the underlying physics. \\
A closer look at the values in table. 1 and the fig.4, reveals that it is probably not the initial values of energy or helicity of the forced or the decaying cases that determines the intermittency at the final state, but the crucial parameter is perhaps the amount of time that is taken to reach that initial state. To understand this statement, we have to look at the initial values of the energies for cases 3ai, 3bi and 3ci, the final values and the resulting intermittent states, along with the time at which these initial values were obtained. It is clearly seen that for case 3a, the initial state was at $t=6.7$ while for cases 3b and 3c it was at  $t < 1$. One can also see the change in intermittency from case 3ai to case 3af and also cases 3bi and 3ci to 3bf and 3cf in the structure function exponent and flatness curves. It is clear from these plots that case 3a shows a significant change in intermittency, while a similar change in cases 3b and 3c is not as pronounced, although the amount of simulation time for all the three cases is approximately $5$ large eddy turnover times. A way to rephrase this discussion is to say that the longer the system is forced (at small scales) before its eventual decay, the larger the change in its intermittency will be, although the forced system by itself may not be significantly intermittent. This could be also be understood as : if the system is helically forced for a shorter amount of time, all the scales of the system probably do not get twisted or linked enough (i.e. do not posses enough magnetic helicity) so as to snap vigorously, once this forcing is removed, and thus show relatively smaller amounts of intermittency. On the other hand, if the forcing exits for long enough time, it is possible that a larger number of structures (and scales) obtain significant magnetic helicity (though its magnitude might be small) and when the forcing is removed, show highly intermittent behavior. \\
The inferences that could be drawn from the above discussion for the observations are (a) it is possible that the topological changes at small scales in ARs are necessary for a flaring process (a conclusion reached in \cite{sor04}) and (b) the larger the amount of time spent creating an initial state in an AR on the photosphere the larger the change in intermittency will be once it flares up. It is necessary to note that this initial state (created over a significantly larger amount of time) may not by itself have any large amount of intermittency (as can be observed from fig. 2b (red curve case 3ai) but is bound to create a state which is highly intermittent (dark green curve case 3af).\\
A significant lacuna in these simulations is that they do not establish what causes an AR to flare up. In all the simulated cases, it is the human intervention that stopped a simulation at a particular instant (in the forced case) and set the system to decay, so as to observe the changes in intermittency. However, in ARs, this process occurs naturally by itself. It appears that currently there are no simulations that determine why and how an AR will snap into a flare. Since we observe several classes of flares emerging from ARs, we can at the most comment on the initial and final states using the  changes in its intermittency once the event has taken place, but at the current moment we lack the power to predict which AR will flare up and when using this model approach.\\
It is also to be noted that any astrophysical system where there exists an initial helical forcing (e.g. jets originating in a merger of cluster of galaxies) that later gets switched off, turning the system into a decaying MHD flow under the simplest possible assumptions, could be interpreted using this simulation data. However, such data from distant astrophysical systems necessary to perform a similar statistical analysis is currently not available due to the limitations in instrumentation and observational techniques. Hence for now, we have to limit ourselves to use this simulation data in an attempt to improve our understanding of statistical properties of ARs on the Sun.
Matching the simulation and observational data units requires a certain level of arbitrariness due to the spatial homogeneity of the 
numerical system. This can, however, be avoided with regard to the similarity properties of both systems as expressed by the scaling
exponents $\zeta_p$ which allow direct comparison.
 \section{Conclusions}
Structure function exponents of ARs are better modeled using a log-Poisson model with Kolmogorov phenomenology and co-dimension 1 during the flaring phase, while they appear to fit with a model curve of co-dimension $1.5$ for the onset phase of the flare. We also observe a deviation of structure function exponent curves from the onset phase to the phase where the actual flare erupts. \\
We performed direct numerical simulations of homogeneous, incompressible 3D-MHD turbulent flow in three cases namely forced, decaying and initially forced and later allowed to decay. In this study, the third case, with its three sub-cases, is a system that shows similar statistical trends as those observed in the statistical properties of ARs.\\
In the structure function exponent studies, two methods have been studied, namely LSA and ESS, to test which method gives better results for the data. From the plots shown in figures of 1c and 1d, it is clear that LSA gives better error estimates and appears to be a more reliable method than ESS, although the information obtained from both these methods is similar. \\
From the structure function plots derived from the simulation data, we can observe an increase in concavity from case 3ci to case 3af, which is a similar trend to that observed in fig.4 of \cite{abr02} for different ARs. This increase in concavity was linked to the increase in flare activity by \cite{abr05}. The multi-fractal nature observed in ARs is also replicated in these structure function exponent plots.\\
Flatness curves of the simulated cases (fig.2c) imitate the different type of ARs described in \cite{abr10}, with the intermittency increasing from forced cases to decaying cases. The correlation function of the final state of all of the studied cases falls to the right of the corresponding initial state, a behavior also noticed in ARs.\\
The change in topology of the structures from the initial forced state to the final decaying state has a significant influence on the observed statistical behavior, with a clear change in the spectral power law of magnetic helicity from $-3.3$ (initial forced state e.g. case 3ai) to $-3.6$ (final decaying state e.g. case 3af). Even then, one cannot completely explain the observed statistical behavior of ARs, simply by evoking the change in topology ({\it{aka}} magnetic helicity) or the change in the value of the energies (both magnetic and kinetic) from the initially forced state(s) to final decaying state(s). However, the most significant inference coming from the simulations that could help us understand the observed behavior of ARs, is the importance of the amount of time the system needs to reach the pre-flaring stage (we call these: `initial state(s)' for each of the subcases of case 3). If the initial state was achieved quickly then the change in intermittency from this state to the final flaring state is not so high. On the other hand, if the pre-flaring state is obtained after a prolonged lapse of time, possibly through an ever present small scale forcing, then the change in intermittency from this state to the final flaring state is much higher.  At the present moment the simulations fail to make any comment on why these so-called initial states flare up at all. For now, to understand the statistical properties of ARs, the pre-flaring state can be treated as some form of  incompressible, homogeneous, forced 3D-MHD flow, where the influence of complex phenomenon like convection from the solar interior and the differential rotation etc..of the Sun are embedded in the small scale forcing term(s) $\bf{F_v},\bf{ F_b}$ of the equations, supplying a helical forcing to the system. In the same manner, the flaring state could be treated as a  incompressible, homogeneous 3D-MHD flow, where due to currently unknown reasons, the drive is switched off and the system behaves as a decaying flow.\\
On the whole, the statistical properties of ARs could be modeled using DNS data, with perhaps the simplest of the MHD turbulent flows making it a good starting point for many such future attempts.
\section{Acknowledgments}
 SKM wants to acknowledge Prof. D. Hughes and Prof. S. Tobias and Dr. C. Davies at University of Leeds, Leeds, UK for their encouragement, Prof. V.I. Abramenko, Big Bear Solar Observatory, CA, USA, for allowing her data to be replotted in fig. 1b and also providing the figures 1a, 2a, 2b and 3a to be reproduced in this work, Prof. A. Pouquet and Prof. P. Mininni at NCAR, Boulder, CO, USA and  Prof. R. Pandit of IISC,Bangalore, India, Dr. D. Mitra, NORDITA, Stockholm, Sweden and Dr. G. Sahoo, MPI Solar system physics, Gottingen, Germany, for useful discussion on various parts of this work.

\clearpage
\begin{figure}[h]
(a)\includegraphics[width=7cm,height=7cm,viewport=-1 1 220 309,clip]{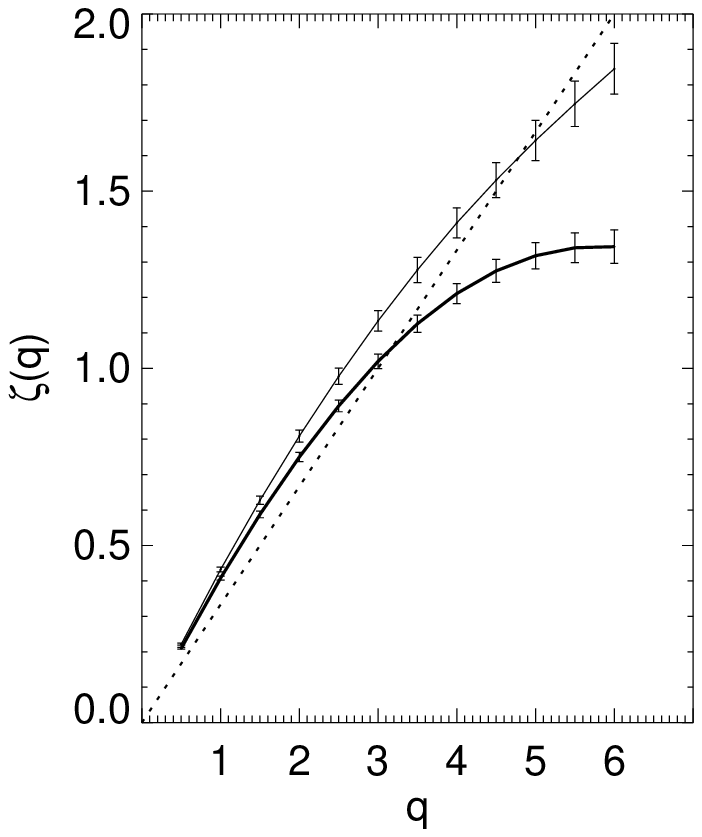}
(b)\includegraphics[width=7cm,height=7cm,viewport=54 14 406 330,clip]{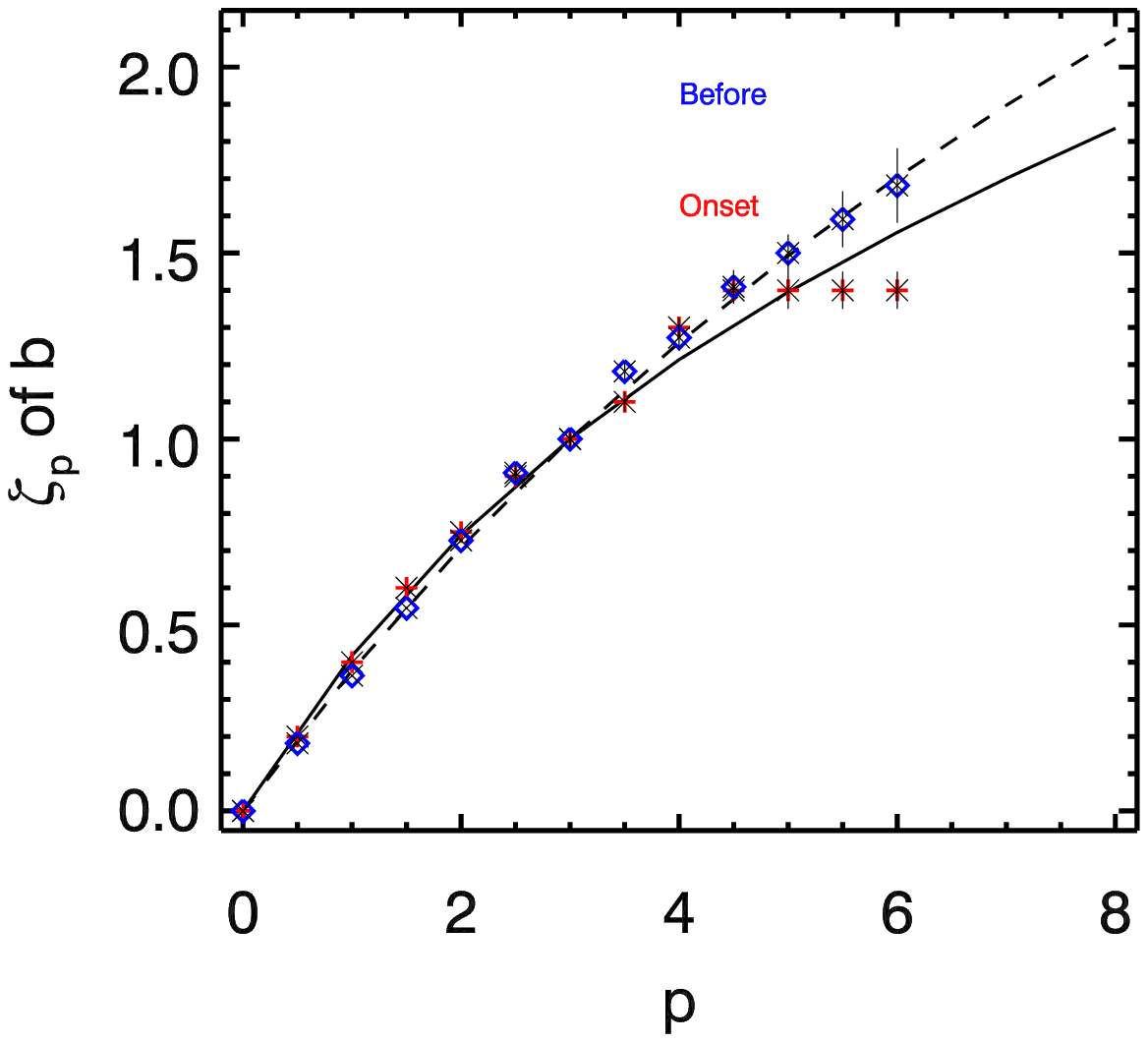}
(c)\includegraphics[width=7cm,height=7cm,viewport=54 14 406 330,clip]{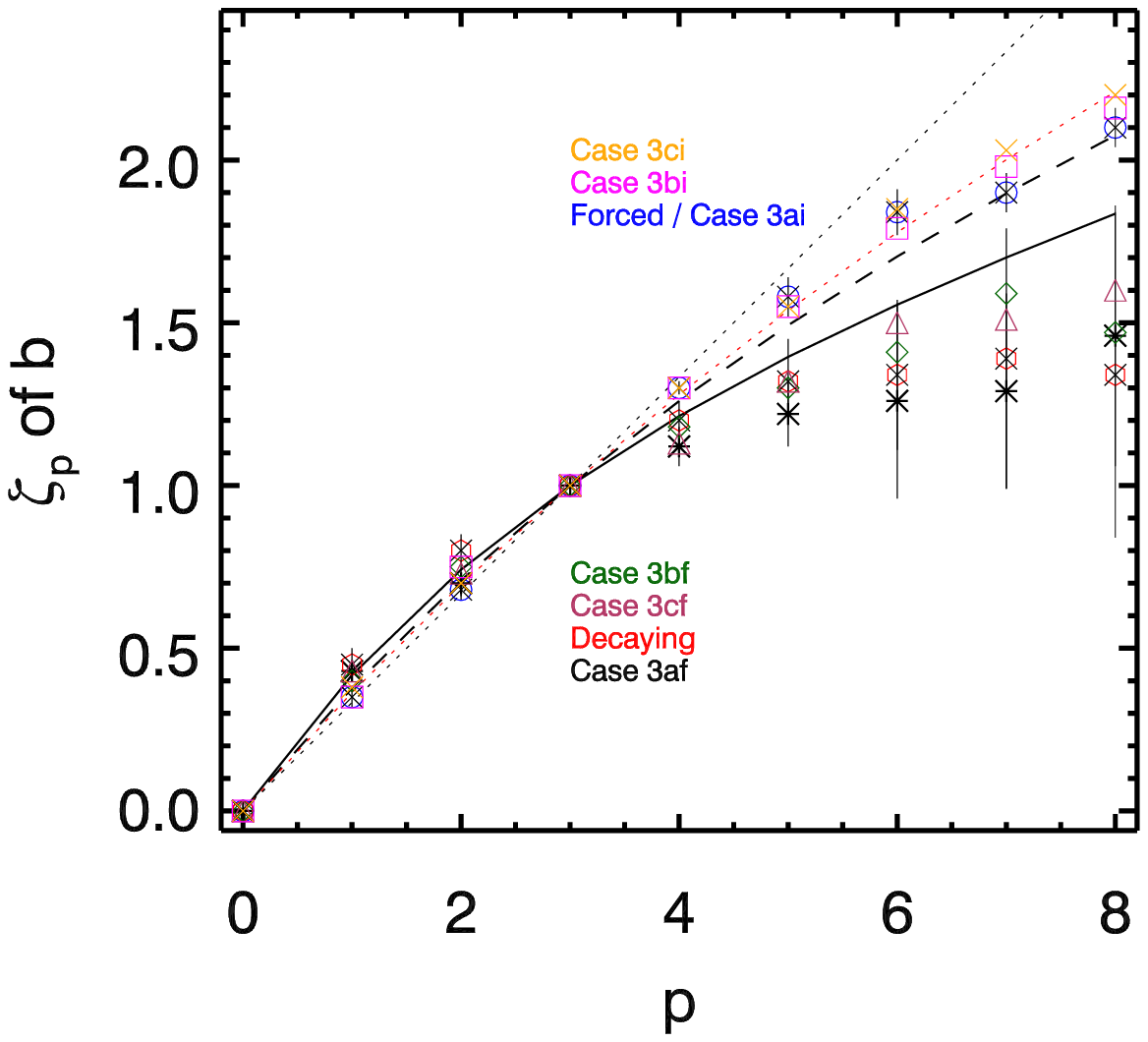}
(d)\includegraphics[width=7cm,height=7cm,viewport=54 14 406 330,clip]{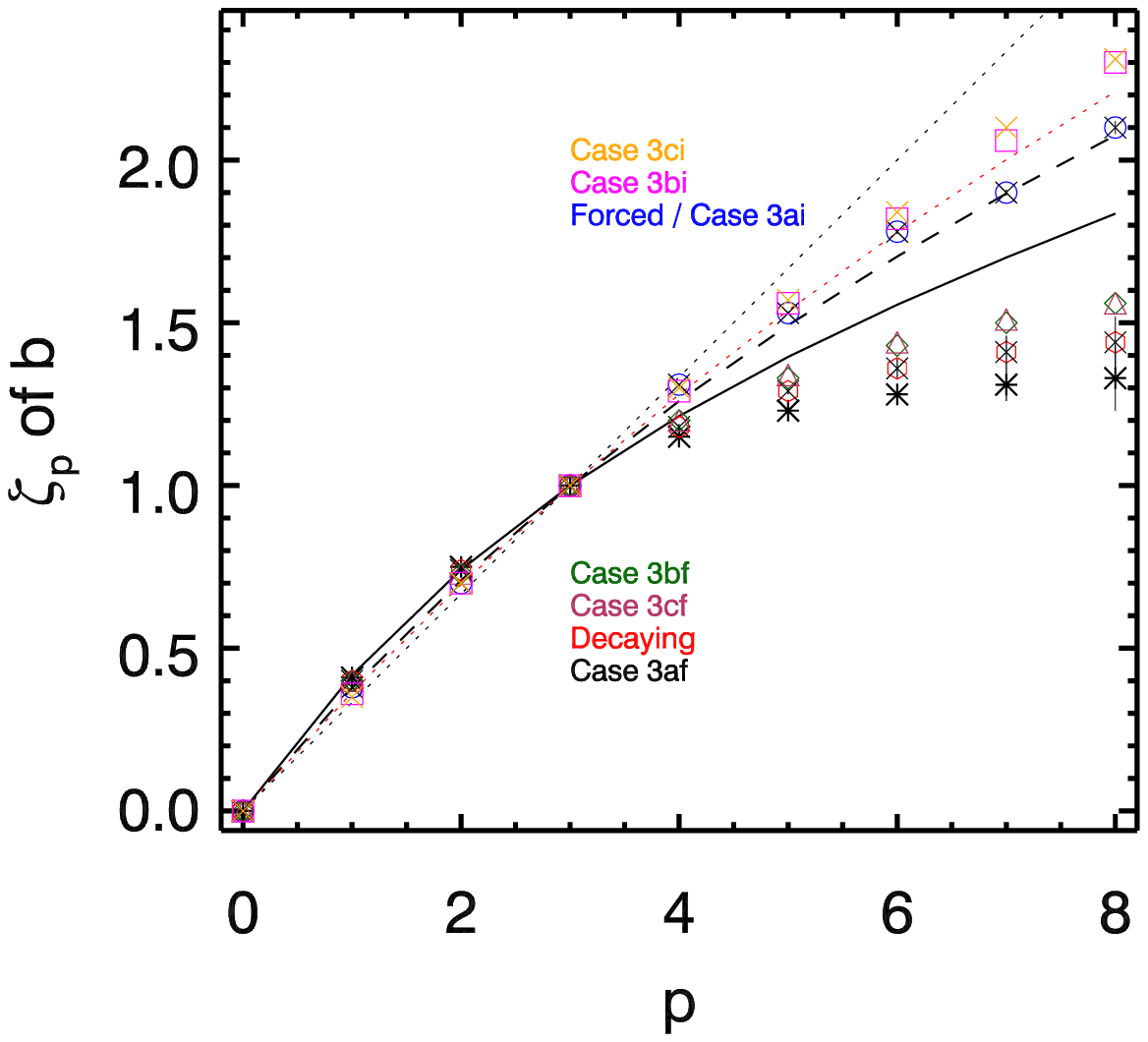}
\caption{ {\small{Structure function exponents ($\zeta_p$) plotted against the order ($\it{p}$) of the structure functions. (a) Reproduction of Figure 2 of \cite{abr03} (courtesy V.I. Abramenko). The dotted line is the K41 non-intermittent model curve. The top curve shows $\zeta_p$ before the onset of the flare and the bottom curve shows $\zeta_p$ during the rising phase of the flare NOAA AR039. (b) Data extracted from fig.1a, replotted along with the log-Poisson model curves. Solid black line: log-Poisson model curve with $C_0 = 1$, dashed black line: log-Poisson model curve with $C_0 = 1.5$. Blue data points: $\zeta_p$ values in the onset phase of the flare, red data points: $\zeta_p$ values during the rising phase, for NOAA AR039. (c) and (d) $\zeta_p$ Vs $p$ plots obtained from simulations. In both the plots the thin black dotted line (top): the K41 non-intermittent model curve and the thin red dotted line(second line from top) is the  log-Poisson model curve with $C_0 = 2$ and the thick black dashed line and thick black solid line are same model curves that are shown in fig.1b. All the  model curves of the log-Poisson model, shown here have been drawn using the Kolmogorov Phenomenology.  In these plots, each case is represented by a color and symbol, which are listed next in the order `case: color and symbol'. case 3ci: orange $\times$, case 3bi: magenta $\square$, Forced / case 3ai : blue $\circ$, case 3bf: dark green $\diamond$, case 3cf : maroon $\vartriangle$, Decaying case: red $\hexagon$ and case 3af: black $*$. Fig. 1c represents the plots drawn using local slope analysis (LSA) and Fig. 1d represents the plots drawn using extended self similarity (ESS) analysis. Errors for cases 3bi, 3ci, 3bf, 3cf are not shown in these figures. [Color version of the figure is available  online]}}}
\label{fig1}
\end{figure}
\begin{figure}[h]
\begin{center}
(a) \includegraphics[width=6cm,height=6cm,viewport=0 19 358 570,clip]{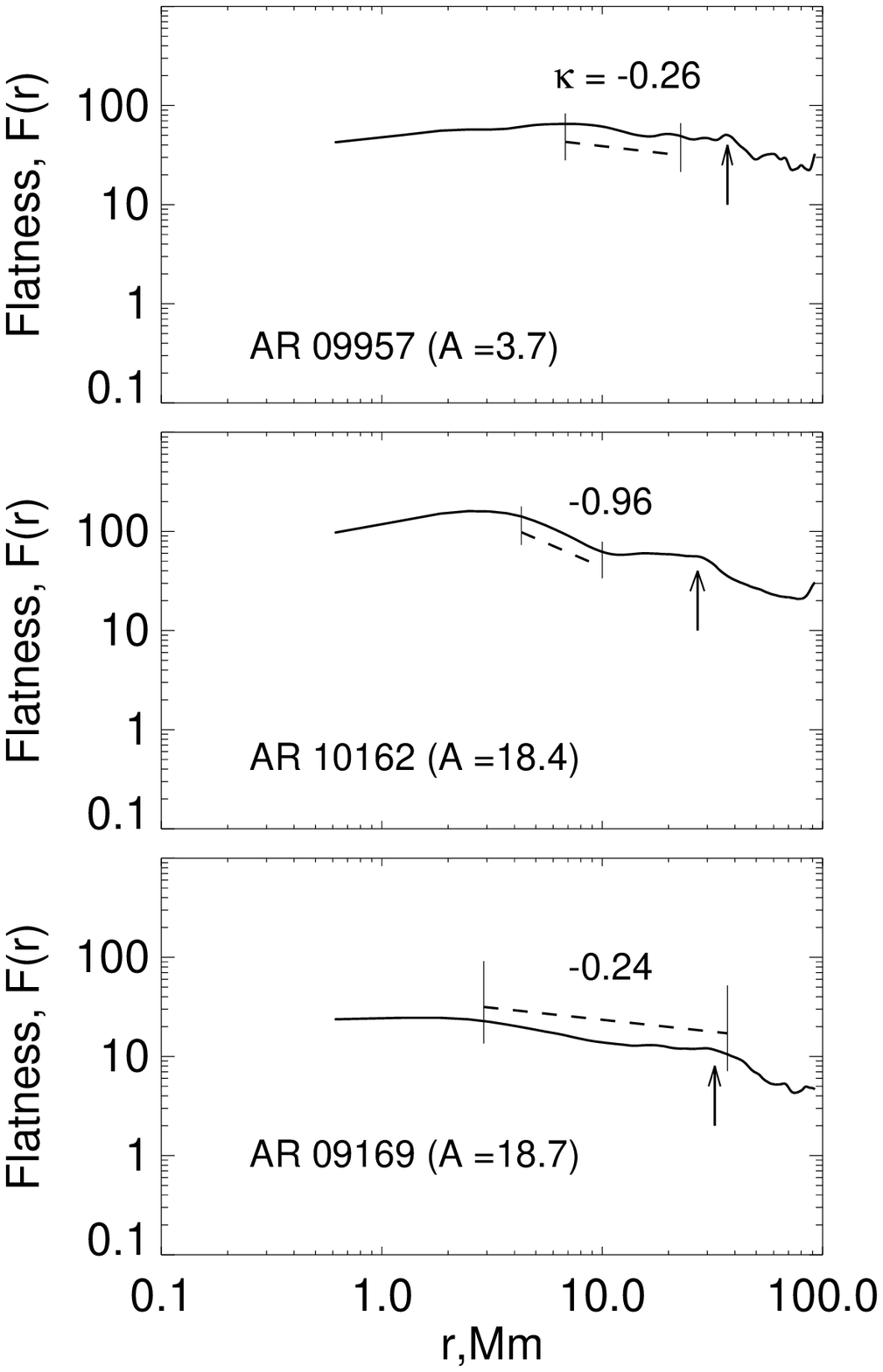}
(b) \includegraphics[width=6cm,height=6cm,viewport=1 8 557 502,clip]{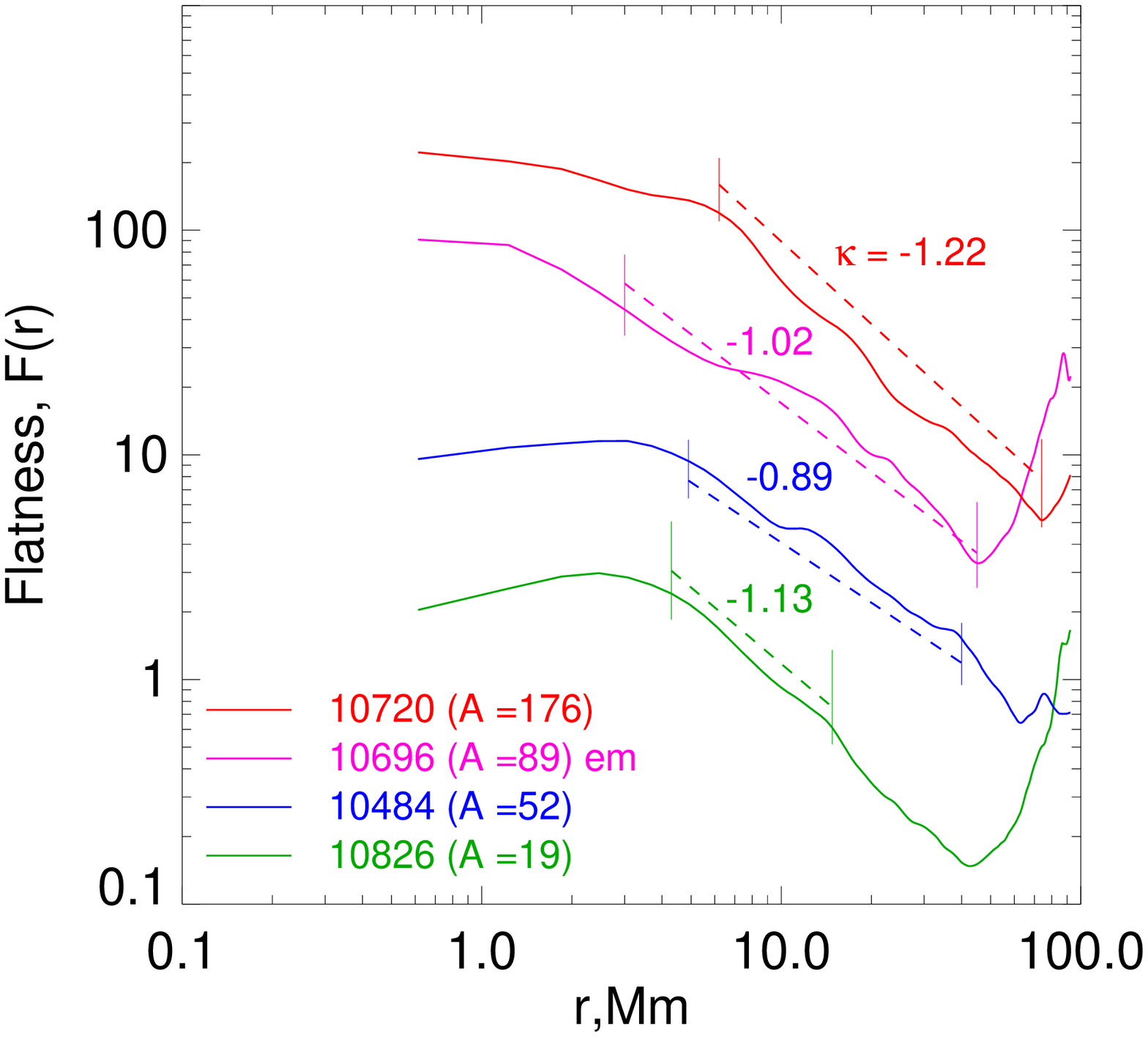}\\

(c) \includegraphics[width=7cm,height=7cm,viewport=54 14 406 330,clip]{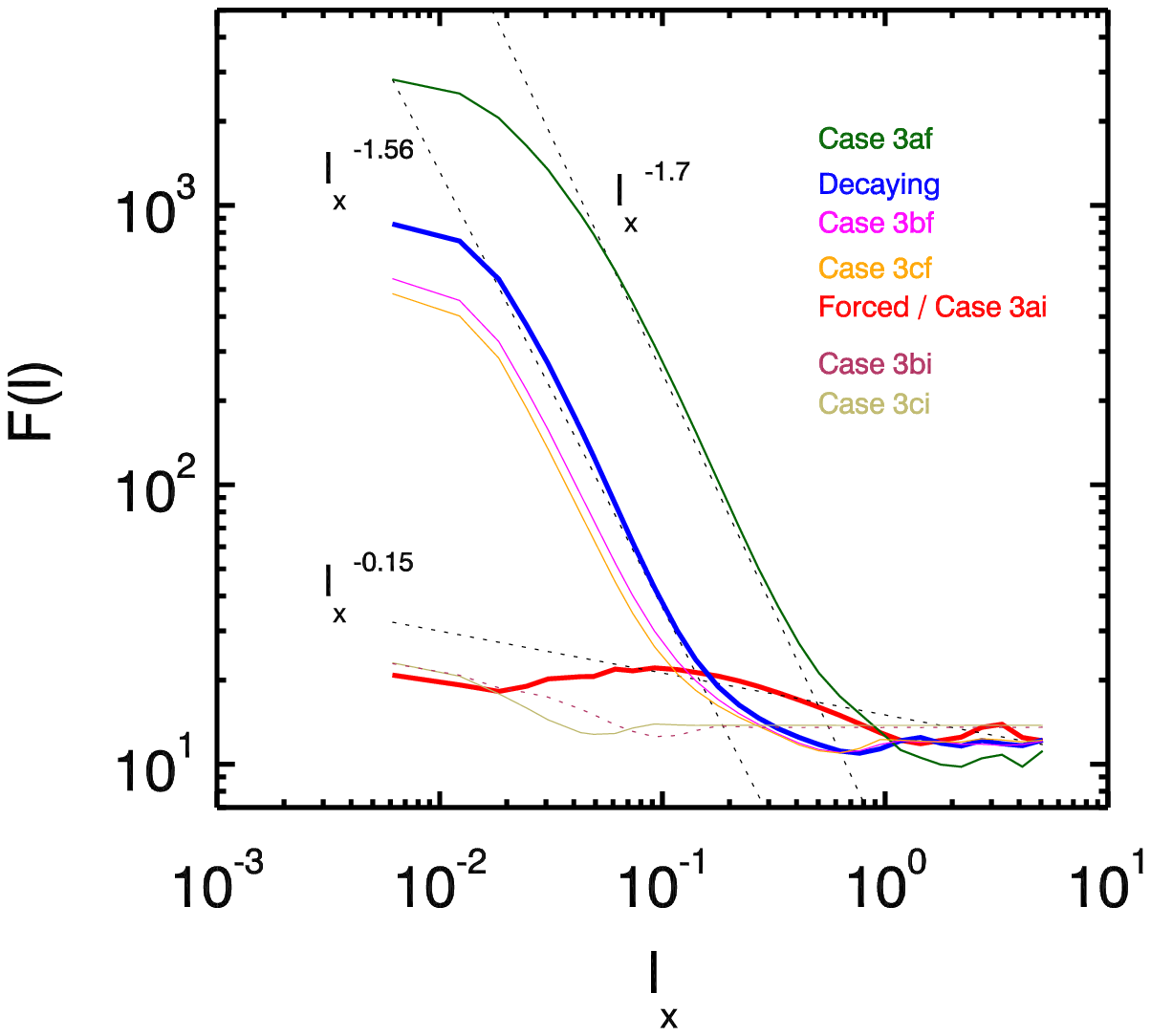}
\end{center}
\caption{Flatness F($\ell$) plotted against the scale length ($\ell$). Figures (a) and (b) are reproduction of figs.4b and 5b of \cite{abr10} (courtesy V.I. Abramenko). Flatness curves for several ARs along with the flatness exponents $\kappa$ are shown in this figure. (c) Flatness curves for the simulated cases are listed in the order `case: type and color of line'(from top to bottom). Case 3af: thick dark green solid line, Decaying case: thick blue solid line, case 3bf: thin magenta solid line, case 3cf: thin orange solid line, Forced / case 3ai: thick red solid line, case 3bi: thin maroon dashed line and case 3ci: thin dark khaki solid line. The three thin black dashed lines represent linear fits to cases 3af, Decaying and Forced / case 3ai, which yield $\kappa$ values of -1.7 (right), -1.56 (middle) and -0.15 (bottom), respectively. [Color version of the figure is available  online]}
\label{fig2}
\end{figure}
\begin{figure}[h]
a) \includegraphics[width=7cm,height=7cm,viewport=7 25 312 312,clip]{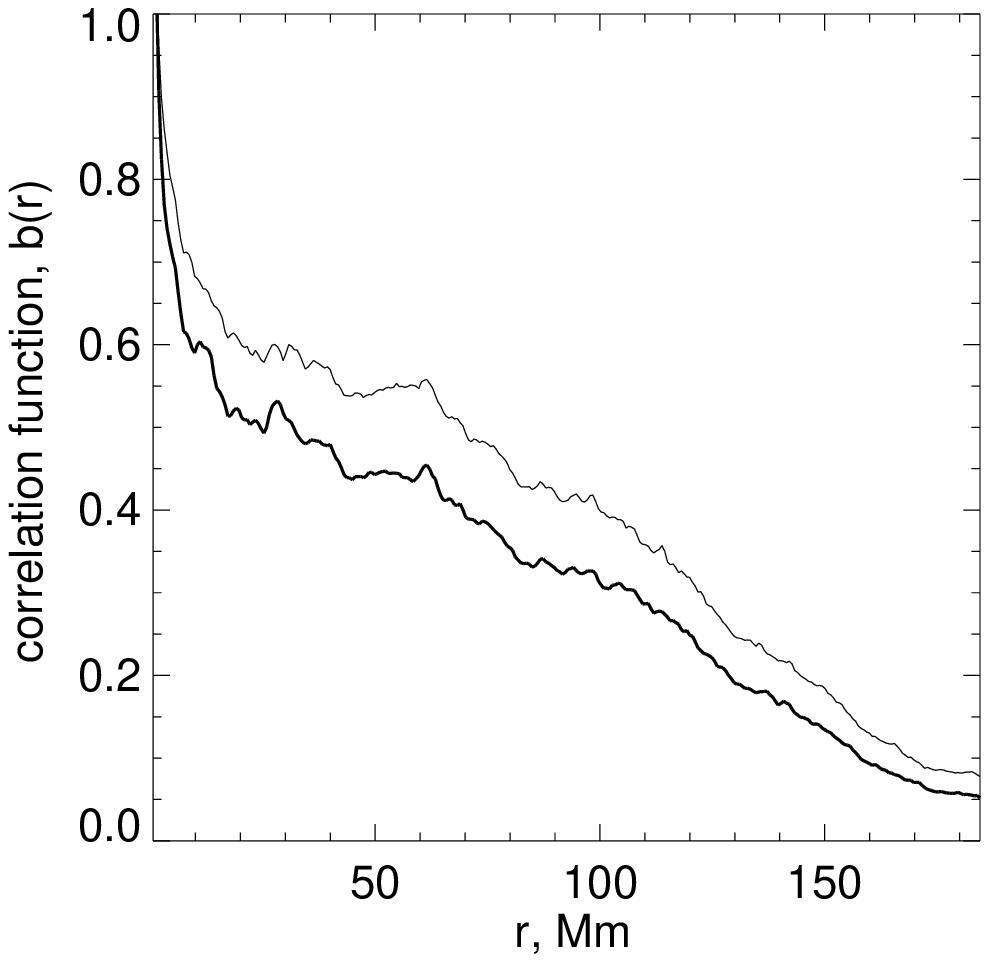}
b) \includegraphics[width=7cm,height=7cm,viewport=54 14 406 330,clip]{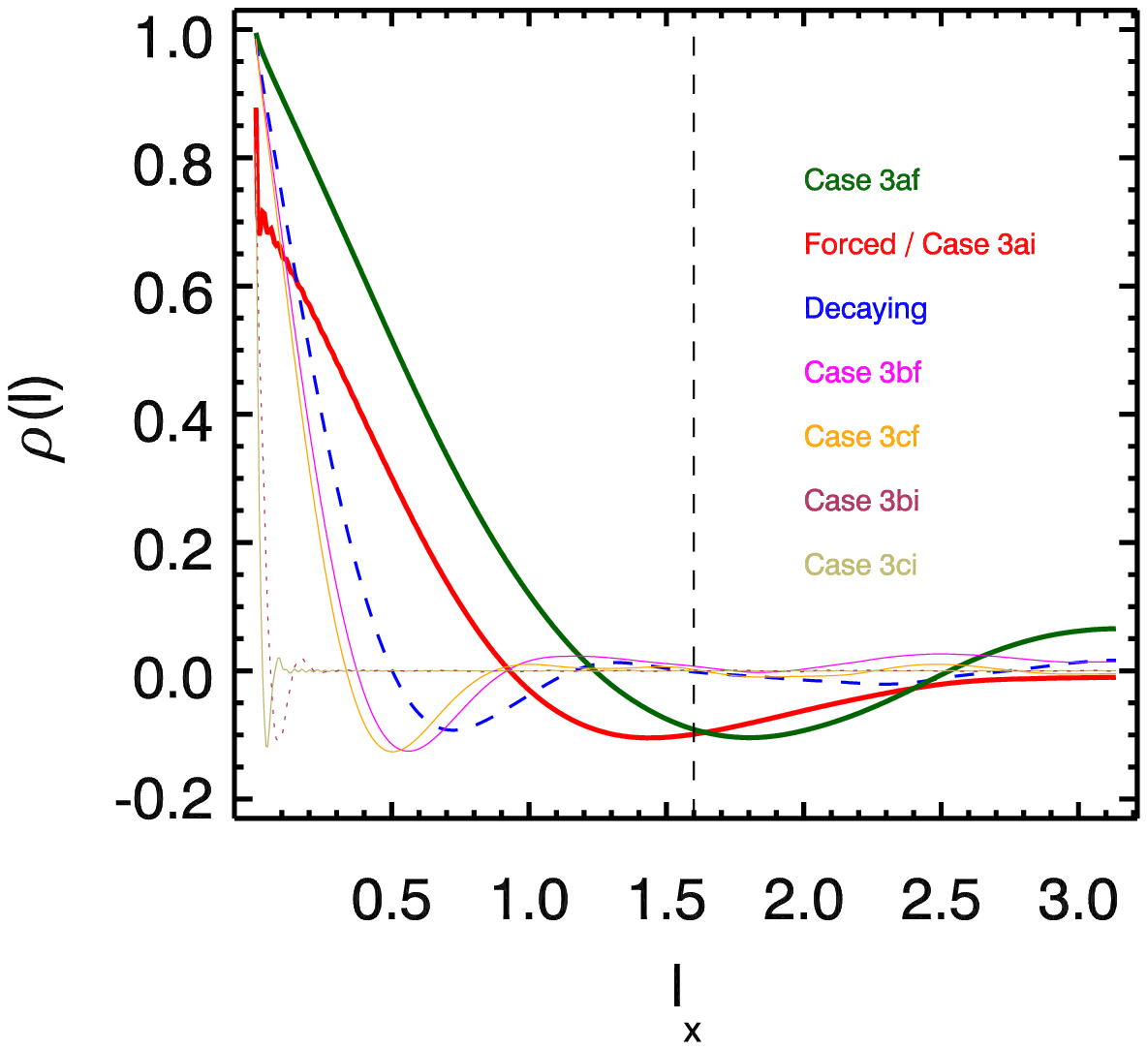}
\caption{Correlation function ($\rho(\ell)$) plotted against scale length ($\ell$). (a) Reproduction of fig.4 from \cite{abr03} (courtesy V.I. Abramenko). Correlation function curves of NOAA AR9661, with lower thick black line representing the phase at the beginning of the flare and thin black line representing the flare near its peak phase. (b) Correlation function curves for the simulated cases are listed in the order `case: type and color of line' (from top to bottom). Case 3af: thick dark green solid line, Decaying case: thick blue solid line, case 3bf: thin magenta solid line, case 3cf: thin orange solid line, Forced / case 3ai: thick red solid line, case 3bi: thin maroon dashed line and case 3ci: thin dark khaki solid line. The vertical dashed line represents the point on the x-axis up to which the trends in fig. 3a are replicated well. [Color version of the figure is available  online]}
\label{fig3}
\end{figure}
\begin{figure}[h]
a) \includegraphics[width=7cm,height=7cm,viewport=35 12 405 400,clip]{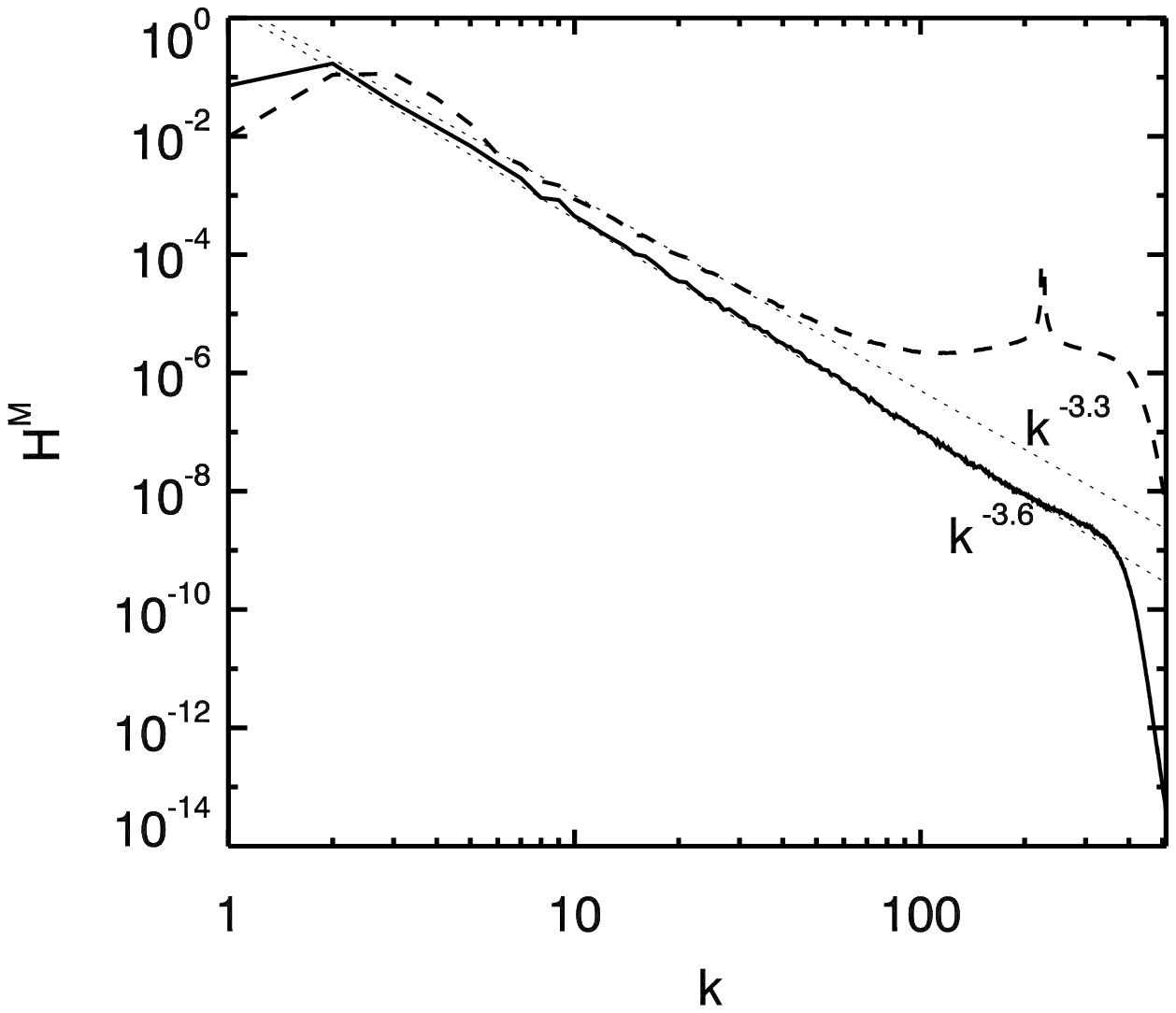}
b) \includegraphics[width=7cm,height=7cm,viewport=35 12 405 400,clip]{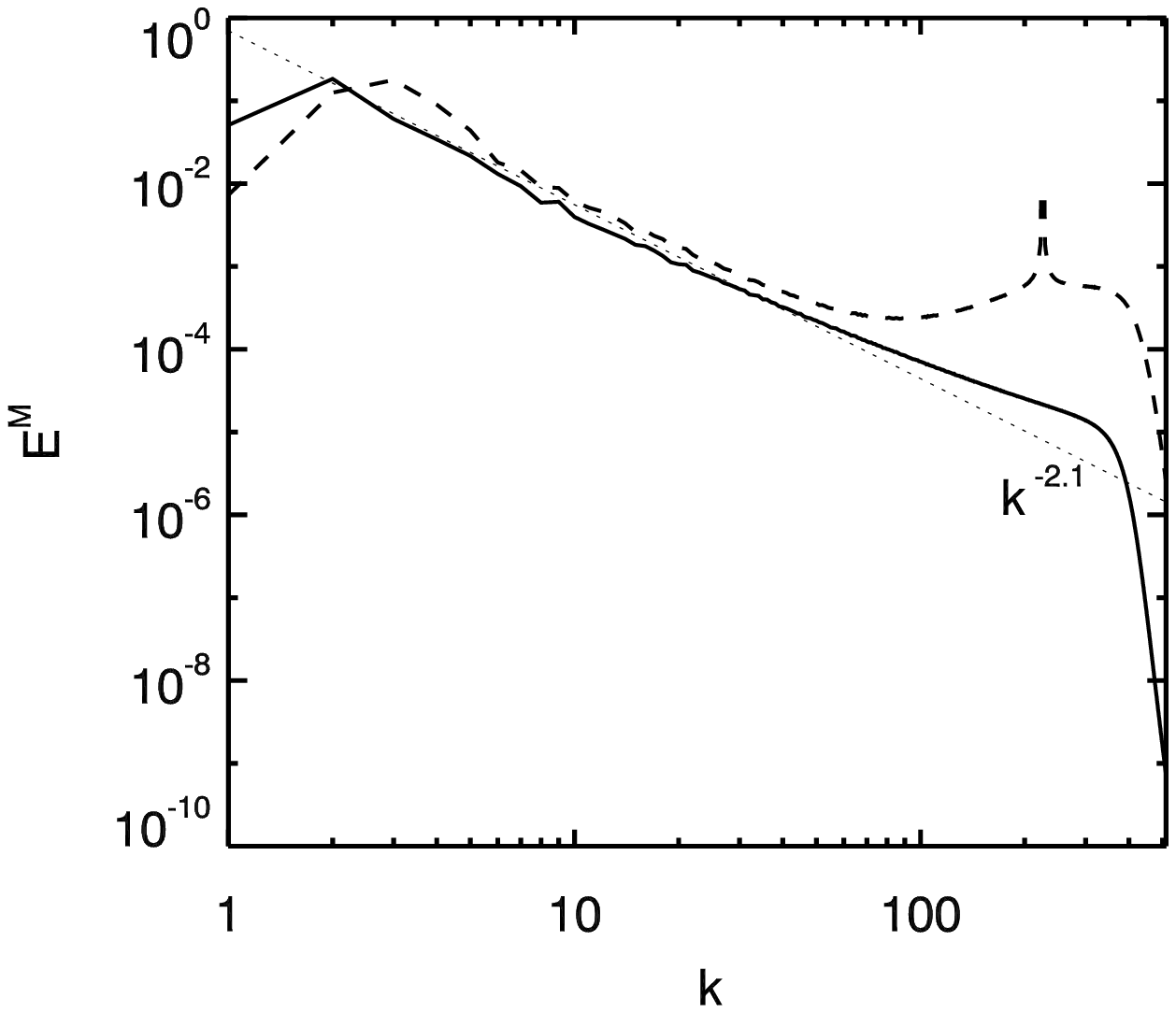}
\caption{Spectral properties of magnetic helicity and magnetic energy for Forced / case 3ai and case 3af. (a) Magnetic helicity $H^M$ Vs Wave number $\it{k}$. Forced / case 3ai: thick black dashed line, case 3af: thick black solid line. Thin black dashed lines represent the power law fits to these plots $k^{-3.3}$ (top) and $k^{-3.6}$ (bottom). (b) Magnetic energy $E^M$ Vs Wave number $\it{k}$. Forced / case 3ai: thick black dashed line, case 3af: thick black solid line. The thin black dashed line represents the power law fit to these plots $k^{-2.1}$.}
\label{fig4} 
\label{fig4}
\end{figure}
\begin{table}[H]
\begin{center}
\small
\begin{tabular}{|c|c|c|c|c|c|c|c|c|c|}
\hline Flow Type $\rightarrow$ & \multicolumn {2}{|c|}{Decaying} &\multicolumn {2}{|c|}{Forced} &\multicolumn {5}{|c|} {case 3} \cr \hline
Measured Value&I&F&I&F /case 3ai& case 3af&case 3bi&case 3bf&case 3ci&case 3f \cr  \hline
$E_{V}$&1&0.0054&0.05&0.1723&0.02873&0.1759&0.008514&0.1480&0.00578\\ 
$E_{M}$ &1&0.02717&0.05&0.6854&0.4353&0.7640&0.1046&0.7947&0.067383 \\
$E_{MR}$&-&$\downarrow$ 37&-&$\uparrow$ 13.7&$\downarrow$ 1.6&-&$\downarrow$ 8.31&-&$\downarrow$ 12 \\ 
$E_{VR}$&-&$\downarrow$ 185&-&$\uparrow$ 3.5&$\downarrow$ 5&-&$\downarrow$ 20.6&-&$\downarrow$ 25 \\ 
$E_{TR}$&-&$\downarrow$ 61&-&$\uparrow$ 9&$\downarrow$ 0.55&-&$\downarrow$ 7.3&-&$\downarrow$ 12 \\ 
$H_{MR}$&-&$\downarrow$ 0.05&-&$\uparrow$ 33&$\downarrow$ 0.01&-&$\downarrow$ 0.03&-&$\downarrow$ 0.05 \\ \hline
\end{tabular}\vspace{2mm}\\
\normalsize
\caption{Values of different physical quantities obtained from the simulation data and their change from the initial (I /i) state to the final (F /f) (See section.4 for more details).}
\end{center}
\end{table} 
\clearpage
\appendix
\section{Structure Functions, ESS \& LSA Figures \& Tables}
 Ordinary structure functions for Forced / case 3ai, Decaying and case 3af are shown in figs. A1, A2 and A3 respectively below. Instead of 6 orders of structure functions plotted for the observations, we plot structure functions up to order 8. The plots for the other cases are not significantly different from these three plots and hence are not shown. In the next step, we use the third order structure function $S_3$ as a reference (on the x-axis) and plot the rest of the structure functions using a logarithmic y-axis, as shown in figs. A4 to A6, corresponding to the figs. A1 to A3. For the other cases the plots are shown in figs. B2, B4, B6 and B8 of the Appendix. To perform ESS analysis, each of the curves in these figures is fitted with a linear fit, yielding a slope that is taken as the structure function exponent $\zeta_p$ at each order. The associated error at each order is calculated using the the standard deviation of several realizations, which is the accepted practice (see e.g. \cite{hol06} and references therein). Local slope analysis of the structure functions is performed by plotting $\frac{dlog10(S_{p})}{dlog10(S_{3})}$ against $log10 S_{3}$ as shown in figs. A7 to A9, B1, B3, B5 and B7. From these plots the region that appears to be mostly linear is selected (indicated by a solid black horizontal line in each of these plots) and the value of the plot on y-axis in that region, is treated as the value of $\zeta_p$ at that order $\it{p}$ (this procedure works mainly for orders up to 4). For higher order plots, where such linearity does not exist the average of all the values under the selected region is taken and is treated as being the $\zeta_p$ at that order. The standard deviation of the rest of the values from this average value is estimated from each of the curves constituting the error at each order (cf. \cite{per09,sah11}). These estimated $\zeta_p$ values from both the methods are shown in tables 3, 4 and 5 along with their respective errors.

\begin{figure}
A1)\includegraphics[width=4.5cm,height=4.5cm,viewport=43 8 410 330,clip]{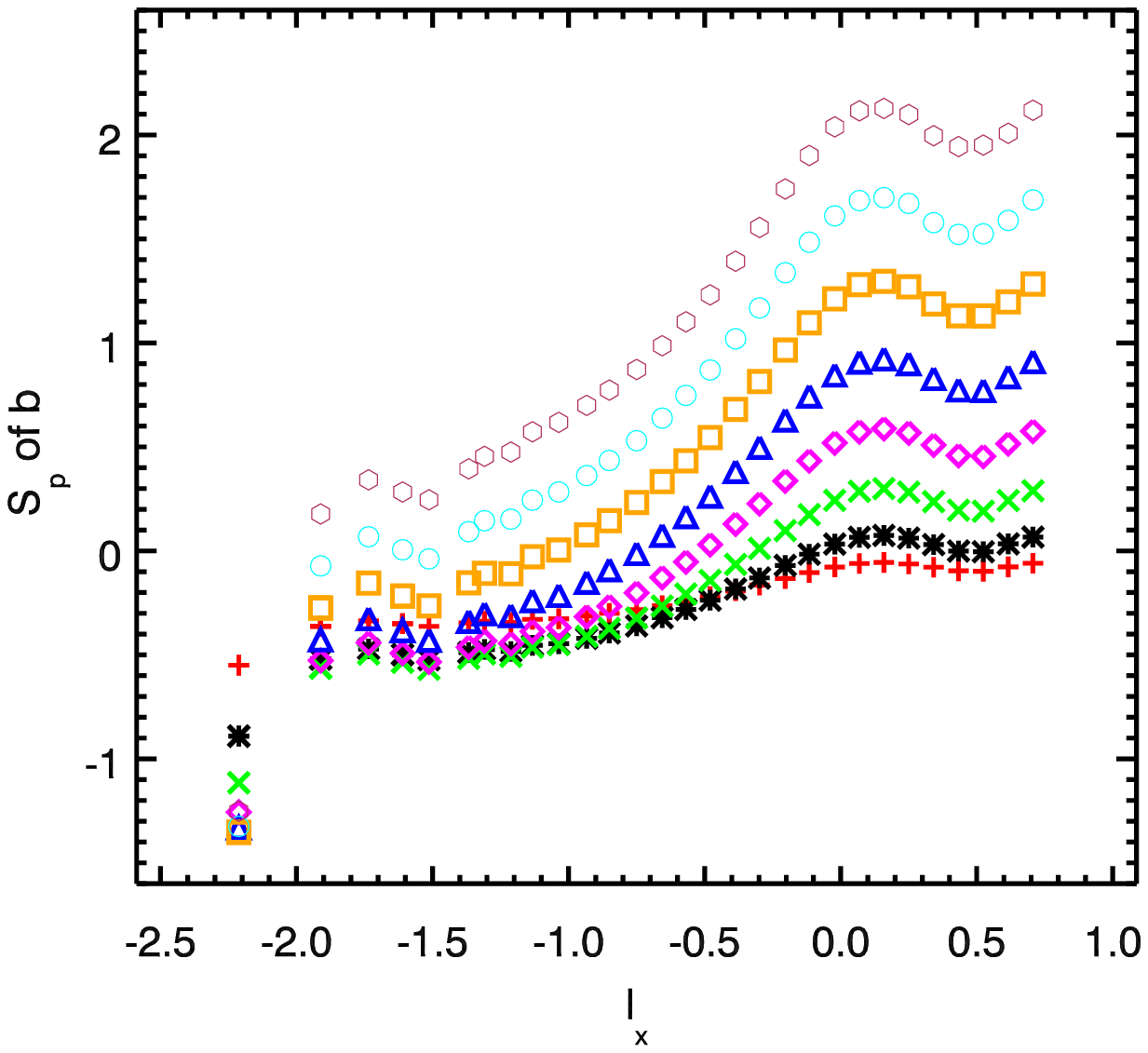}
A2)\includegraphics[width=4.5cm,height=4.5cm,viewport=43 8 410 330,clip]{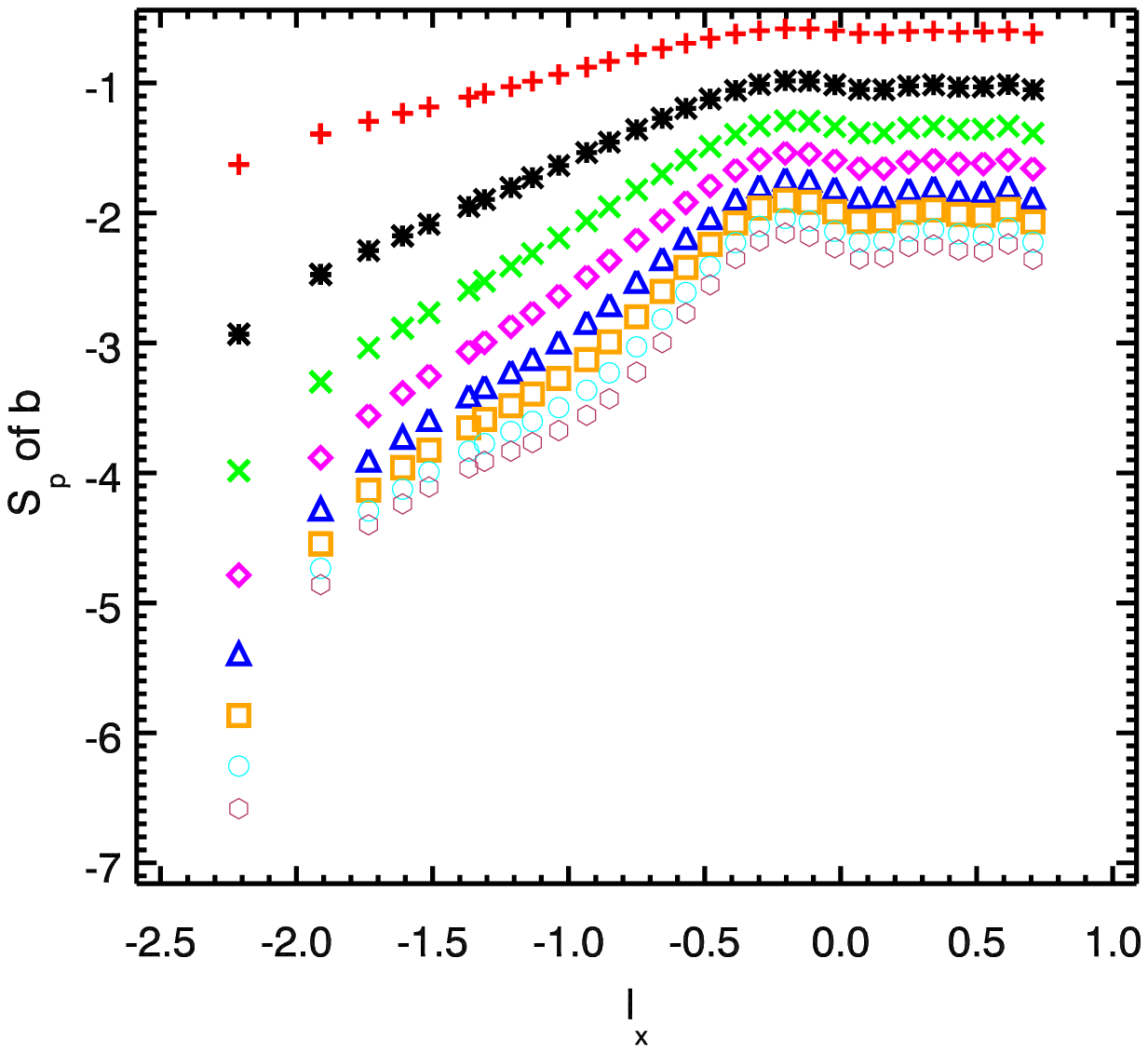}
A3)\includegraphics[width=4.5cm,height=4.5cm,viewport=43 8 410 330,clip]{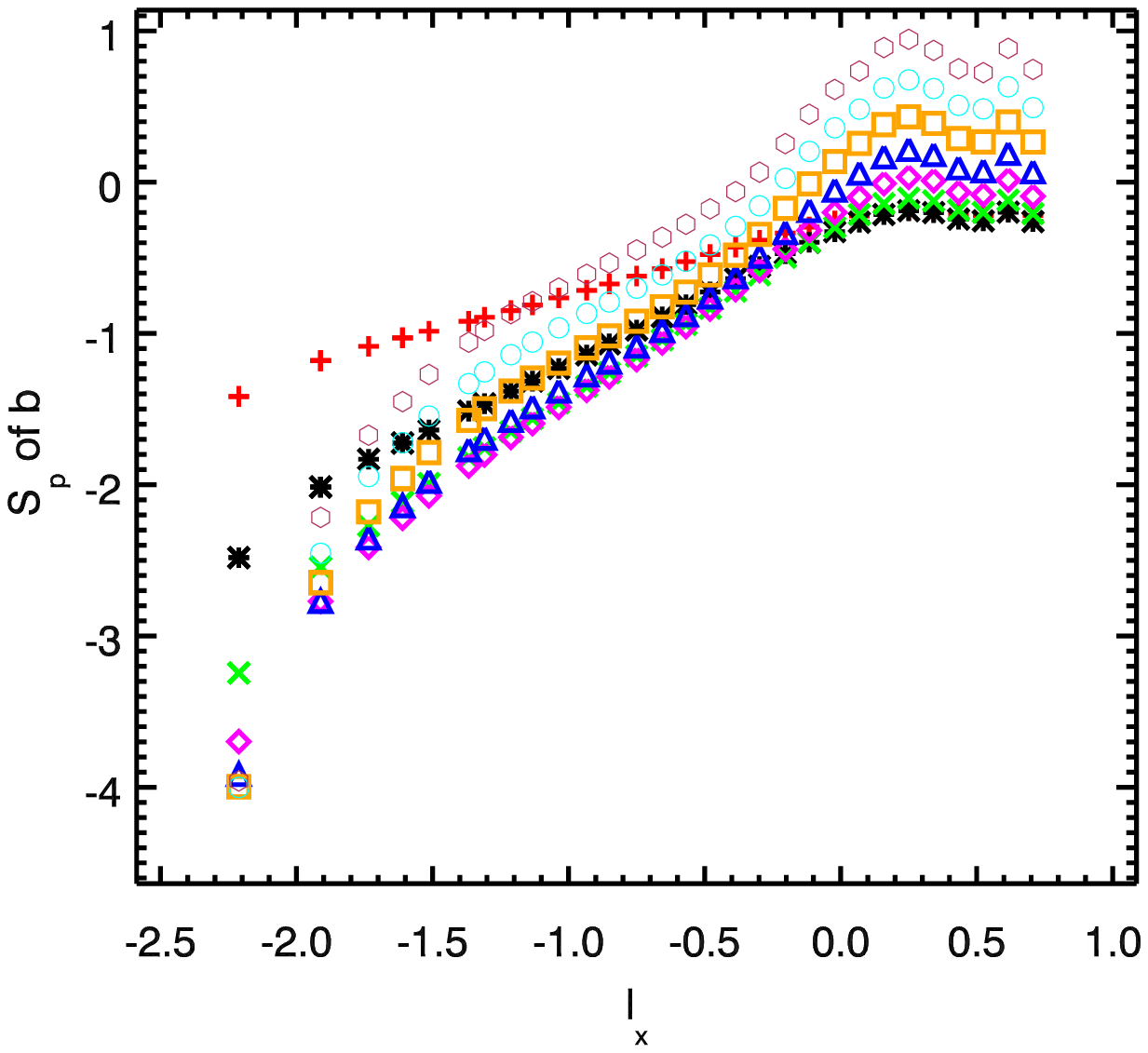}\\
A4)\includegraphics[width=4.5cm,height=4.5cm,viewport=43 8 410 330,clip]{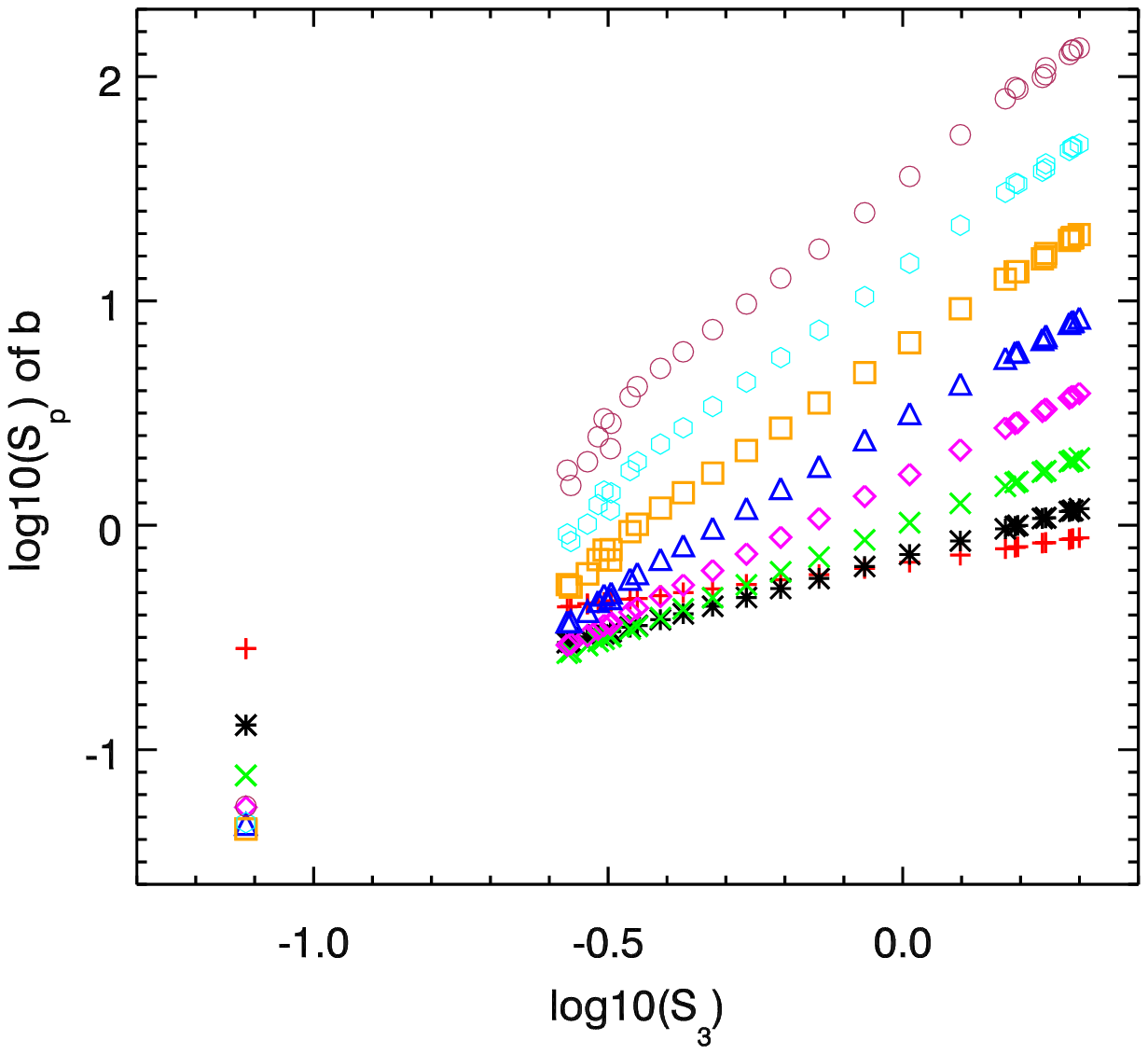}
A5)\includegraphics[width=4.5cm,height=4.5cm,viewport=43 8 410 330,clip]{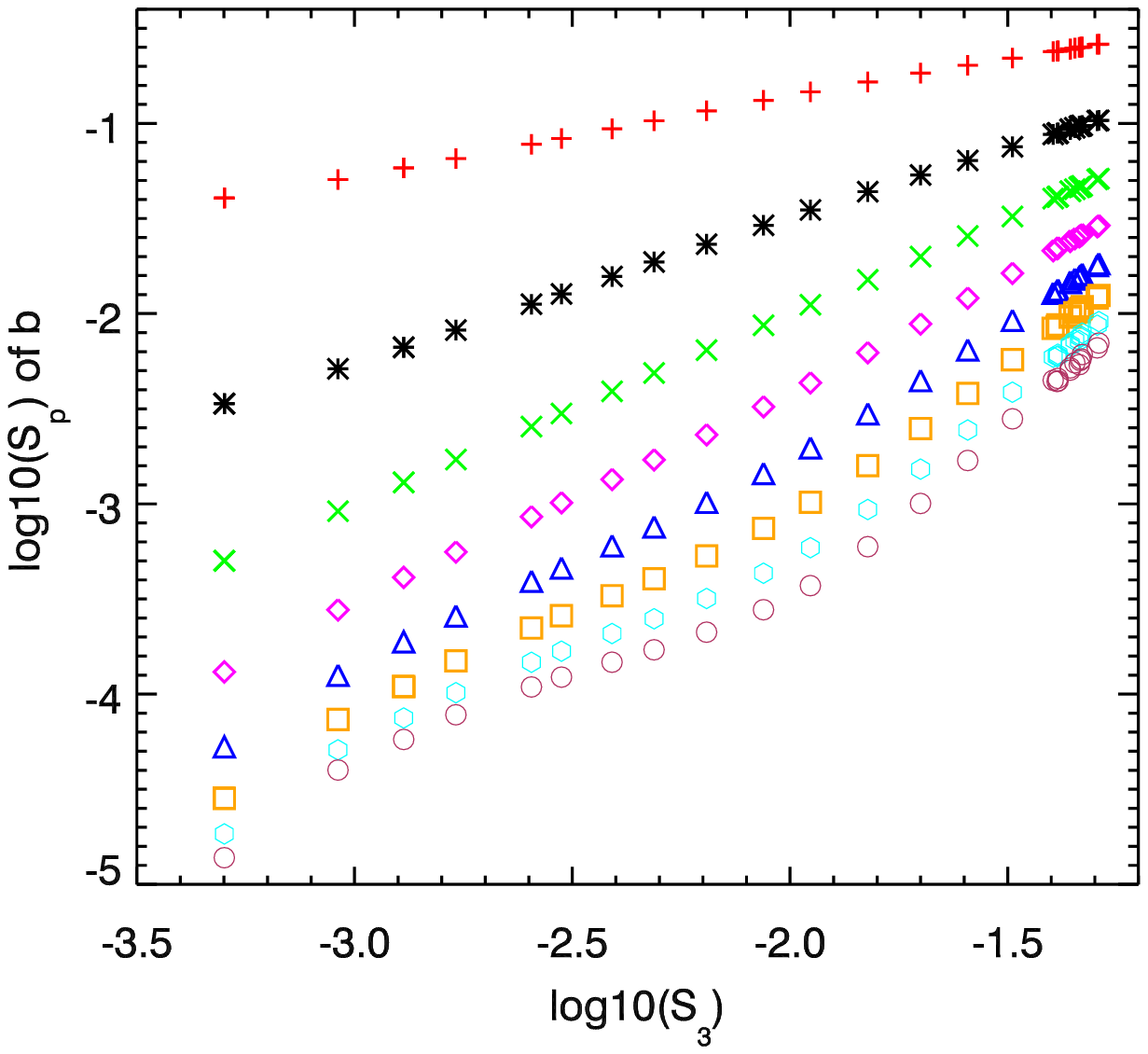} 
A6)\includegraphics[width=4.5cm,height=4.5cm,viewport=43 8 410 330,clip]{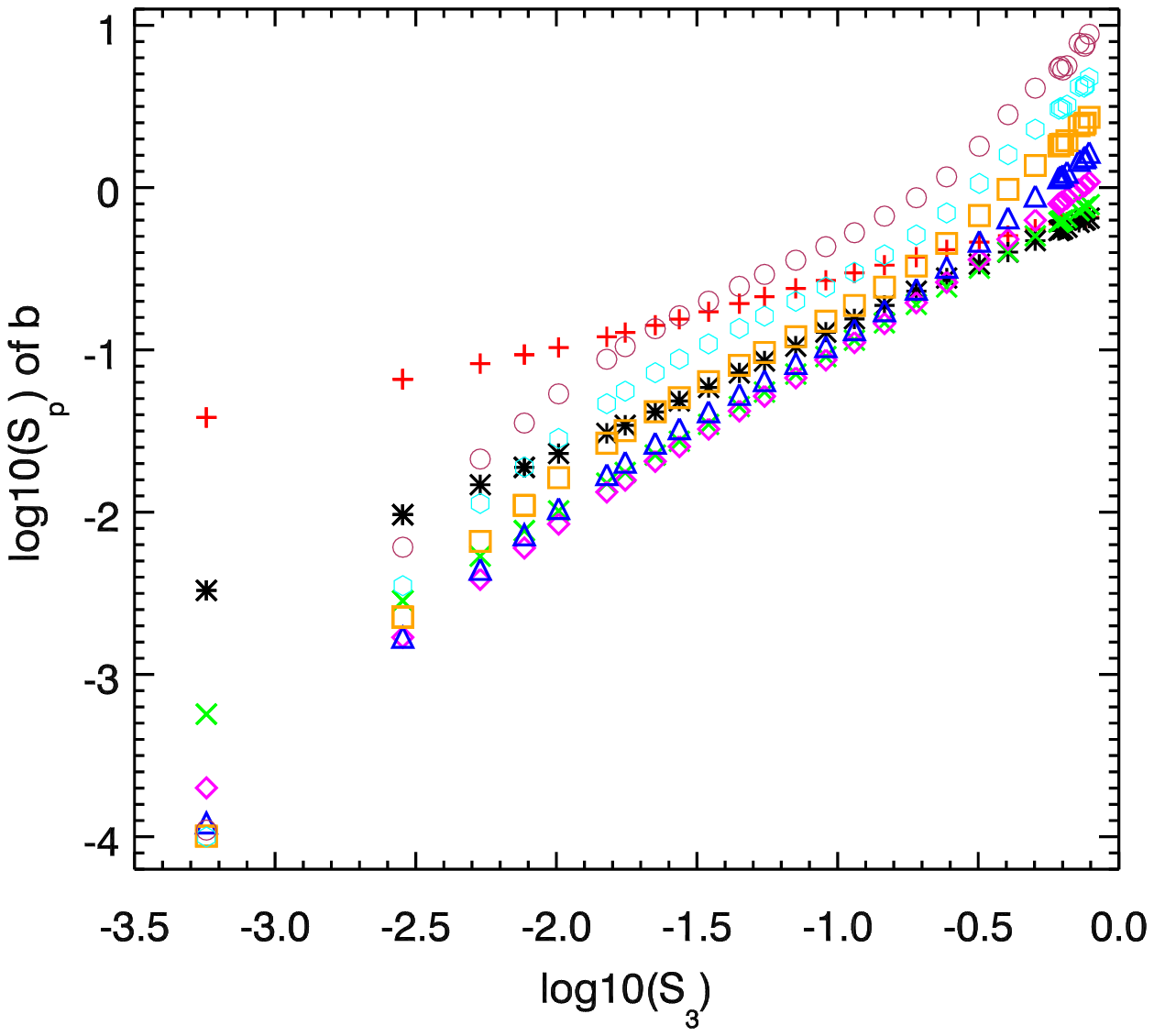}\\
A7)\includegraphics[width=4.5cm,height=4.5cm,viewport=43 8 410 330,clip]{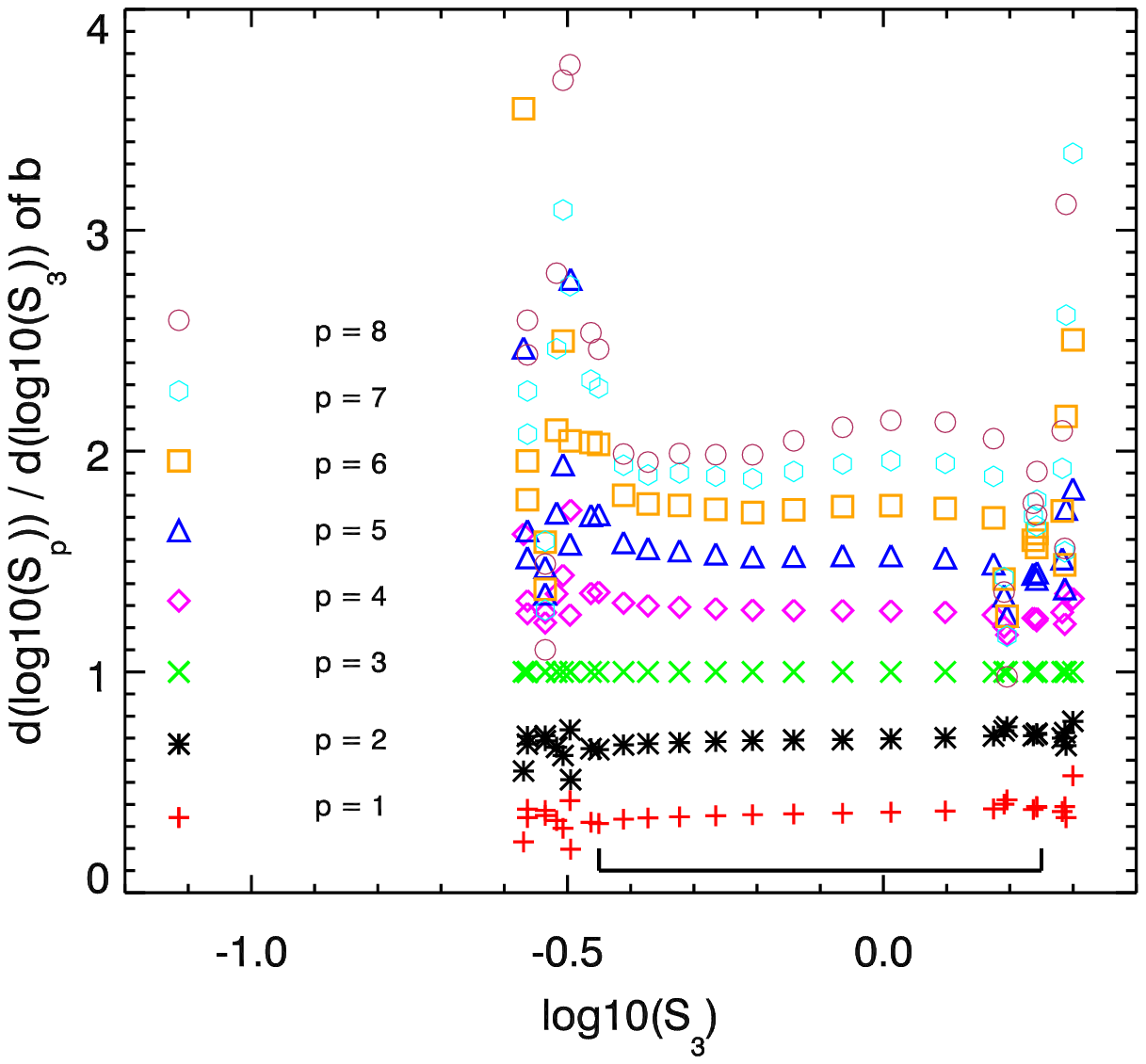}
A8)\includegraphics[width=4.5cm,height=4.5cm,viewport=43 8 410 330,clip]{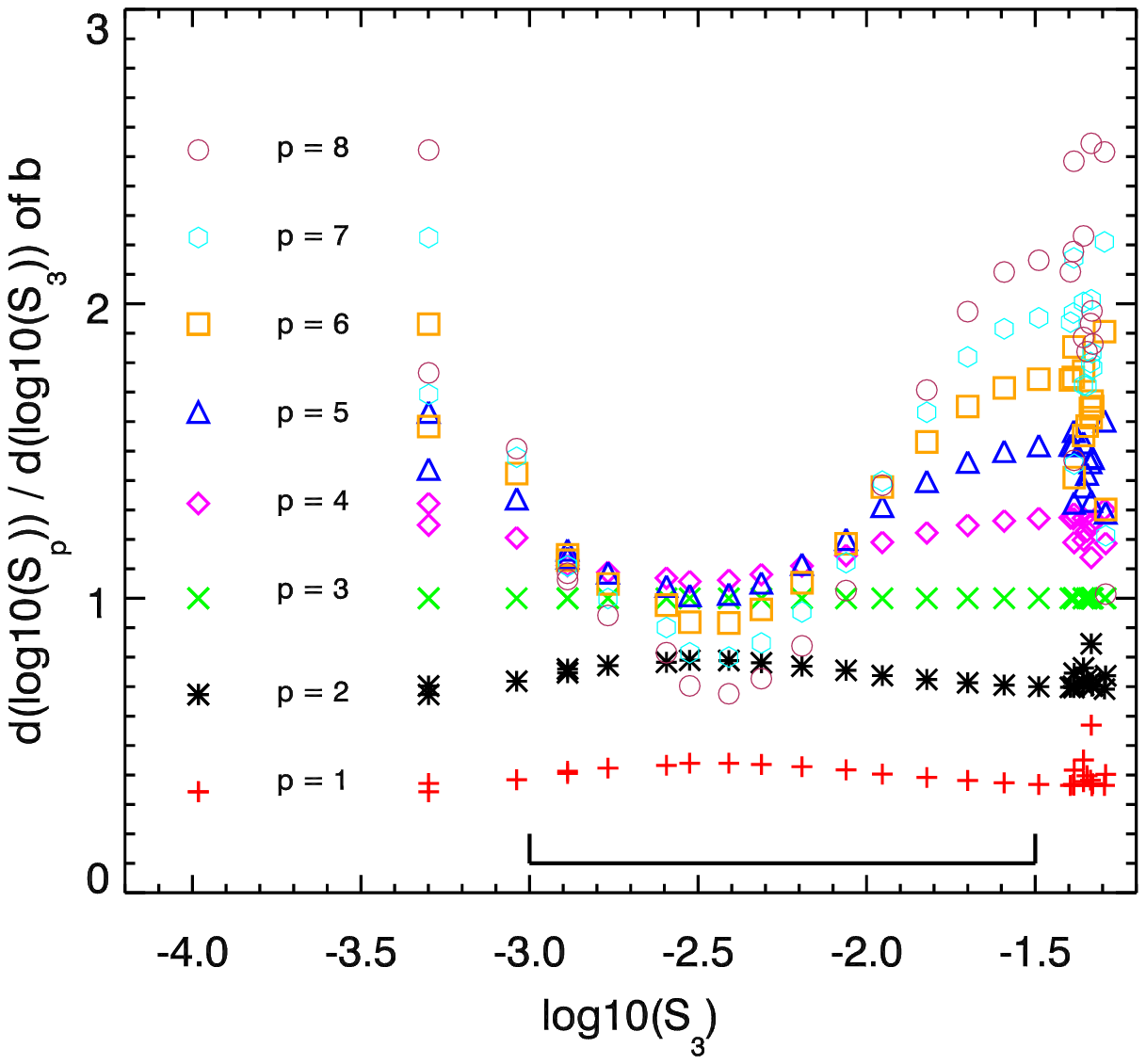}
A9)\includegraphics[width=4.5cm,height=4.5cm,viewport=43 8 410 330,clip]{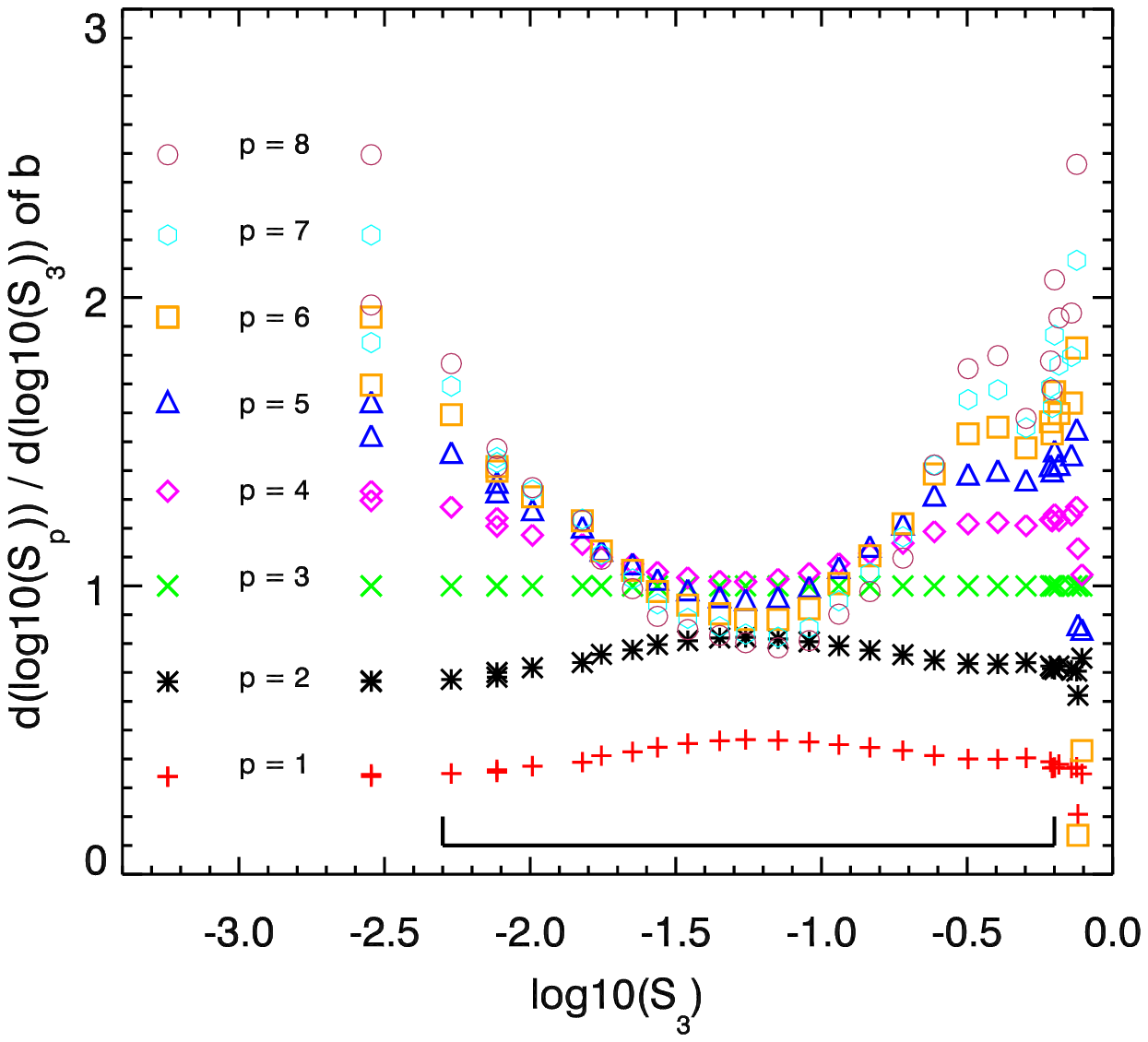}
\caption{The structure functions calculated from simulations for magnetic field for Forced (A1, A4 and A7), Decaying (A2, A5 and A8) and case 3af (A3, A6 and A9) cases. A1 - A3: Ordinary Structure functions, A4 - A6 : log10($S_p$) Vs log10($S_3$) for Extended Self Similarity Analysis (ESS). A7 - A9 : $\frac{dlog10(S_{p})}{dlog10(S_{3})}$ Vs log10($S_3$) for Local Slope Analysis (LSA). Order($\it{p}$): Color and Symbol: 1: red $+$, 2: black $*$, 3: green $\times$, 4: magenta $\diamond$, 5: blue $\vartriangle$, 6: orange $\square$, 7: cyan $\hexagon$ and 8: maroon $\circ$. [Color version of the figure is available  online]}
\end{figure}
\begin{table}
\begin{center}
\begin{tabular}{|c|c|c|}
\hline
Order $\it{p}$& $\zeta_{p}$ During& $\zeta_{p}$ Before\\ \hline
0&0&0\\ 
0.5&0.2$\pm$0&0.2$\pm$0\\ 
1&0.4$\pm$0&0.4$\pm$0\\ 
1.5&0.6$\pm$0&0.6$\pm$0\\ 
2&0.75$\pm$0&0.8$\pm$0\\
2.5&0.9$\pm$0.015&1.0$\pm$0.025\\
3&1.0$\pm$0.015&1.1$\pm$0.025\\
3.5&1.1$\pm$0.025&1.3$\pm$0.025\\
4&1.3$\pm$0.025&1.4$\pm$0.035\\
4.5&1.4$\pm$0.025&1.55$\pm$0.045\\
5&1.4$\pm$0.05&1.65$\pm$0.05\\
5.5&1.4$\pm$0.05&1.75$\pm$0.075\\
6&1.4$\pm$0.05&1.85$\pm$0.1\\ \hline
\end{tabular}\vspace{2mm}\\
\caption{Structure function exponent values obtained from \cite{abr03} fig.2 along with their estimated errors.} 
\label{tab1}
\end{center}
\end{table}
\begin{table}
\begin{center}
\begin{tabular}{|c|c|c||c|c||c|c|}
\hline
 Flow Type $\rightarrow$ & \multicolumn{2}{c|}{Forced / case 3ai $\zeta_{p}$}& \multicolumn{2}{|c|}{Decay $\zeta_{p}$} & \multicolumn{2}{|c|}{case 3af $\zeta_{p}$} \cr \hline
 order $\it{p}$ &LSA&ESS&LSA&ESS&LSA&ESS \\ \hline
1&0.35$\pm$0.025&0.38$\pm$0.001&0.45$\pm$0.05&0.4$\pm$0.001&0.43$\pm$0.03&0.41$\pm$0.001\\ 
2 &0.68$\pm$ 0.02&0.7$\pm$0.001&0.8$\pm$0.05&0.74$\pm$0.003&0.77$\pm$0.05&0.75$\pm$0.003 \\
3 &1&1&1&1&1&1\\ 
4 &1.3$\pm$0.025&1.31$\pm$0.002&1.2$\pm$0.06&1.18$\pm$0.009&1.12$\pm$0.06&1.15$\pm$0.009\\ 
5 &1.58$\pm$0.06&1.53$\pm$0.005&1.32$\pm$0.13&1.29$\pm$0.01&1.22$\pm$0.1&1.23$\pm$0.01\\ 
6 &1.84$\pm$0.07&1.78$\pm$0.007&1.34$\pm$0.23&1.36$\pm$0.03&1.26$\pm$0.3&1.28$\pm$0.03\\ 
7 &1.9$\pm$0.06&1.9$\pm$0.009&1.39$\pm$0.38&1.41$\pm$0.05&1.29$\pm$0.3&1.31$\pm$0.05\\ 
8 &2.1$\pm$0.06&2.1$\pm$0.02&1.34$\pm$0.48&1.44$\pm$0.08&1.46$\pm$0.4&1.33$\pm$0.1\\ \hline
\end{tabular}\vspace{2mm}\\
\caption{Structure function exponents of magnetic field for the following  cases : Forced / case 3ai, Decaying and case 3af. The values obtained from both local slope analysis (LSA) and extended self similarity (ESS) analysis along with respective errors in each case are shown.}
\end{center}
\end{table}
\begin{table}
\begin{center}
\begin{tabular}{|c|c|c||c|c|}
\hline
 Flow Type $\rightarrow$ & \multicolumn{2}{c|}{case 3bi $\zeta_{p}$}& \multicolumn{2}{|c|}{case 3bf $\zeta_{p}$}\cr \hline
 order $\it{p}$ &LSA&ESS&LSA&ESS \\ \hline
1&0.38$\pm$0.025&0.35$\pm$0.001&0.4$\pm$0.05&0.4$\pm$0.001\\ 
2 &0.7$\pm$ 0.025&0.7$\pm$0.001&0.75$\pm$0.05&0.73$\pm$0.003\\
3 &1&1&1&1\\ 
4 &1.3$\pm$0.05&1.3$\pm$0.005&1.18$\pm$0.08&1.2$\pm$0.009\\ 
5 &1.55$\pm$0.05&1.57$\pm$0.009&1.3$\pm$0.12&1.33$\pm$0.01\\ 
6 &1.85$\pm$0.1&1.84$\pm$0.01&1.41$\pm$0.22&1.43$\pm$0.02\\ 
7 &2.03$\pm$0.13&2.1$\pm$0.02&1.59$\pm$0.32&1.5$\pm$0.03\\ 
8 &2.2$\pm$0.13&2.31$\pm$0.03&1.47$\pm$0.38&1.56$\pm$0.05\\ \hline
\end{tabular}\vspace{2mm}\\
\caption{Structure function exponents of magnetic field for cases : case 3bi and case 3bf. The values obtained from LSA and ESS analysis along with respective errors in each case are shown.}
\end{center}
\end{table}
\begin{table}
\begin{center}
\begin{tabular}{|c|c|c||c|c|}
\hline
 Flow Type $\rightarrow$ & \multicolumn{2}{c|}{case 3ci $\zeta_{p}$}& \multicolumn{2}{|c|}{case 3cf $\zeta_{p}$}\cr \hline
 order $\it{p}$ &LSA&ESS&LSA&ESS \\ \hline
1&0.35$\pm$0.01&0.36$\pm$0.001&0.43$\pm$0.03&0.4$\pm$0.001\\ 
2 &0.75$\pm$ 0.01&0.7$\pm$0.006&0.73$\pm$0.03&0.73$\pm$0.003\\
3 &1&1&1&1\\ 
4 &1.3$\pm$0.01&1.29$\pm$0.009&1.13$\pm$0.04&1.2$\pm$0.009\\ 
5 &1.55$\pm$0.05&1.56$\pm$0.01&1.32$\pm$0.1&1.34$\pm$0.01\\ 
6 &1.79$\pm$0.12&1.82$\pm$0.015&1.5$\pm$0.22&1.44$\pm$0.02\\ 
7 &1.98$\pm$0.17&2.06$\pm$0.02&1.51$\pm$0.36&1.51$\pm$0.03\\ 
8 &2.16$\pm$0.25&2.3$\pm$0.03&1.6$\pm$0.39&1.56$\pm$0.05\\ \hline
\end{tabular}\vspace{2mm}\\
\caption{Structure function exponents of magnetic field for  cases : case 3ci and case 3cf. Th values obtained from LSA and ESS analysis along with respective errors in each case are shown.}
\end{center}
\end{table}
\begin{figure}
\begin{center}
B1)\includegraphics[width=4.5cm,height=4.5cm,viewport=43 8 410 330,clip]{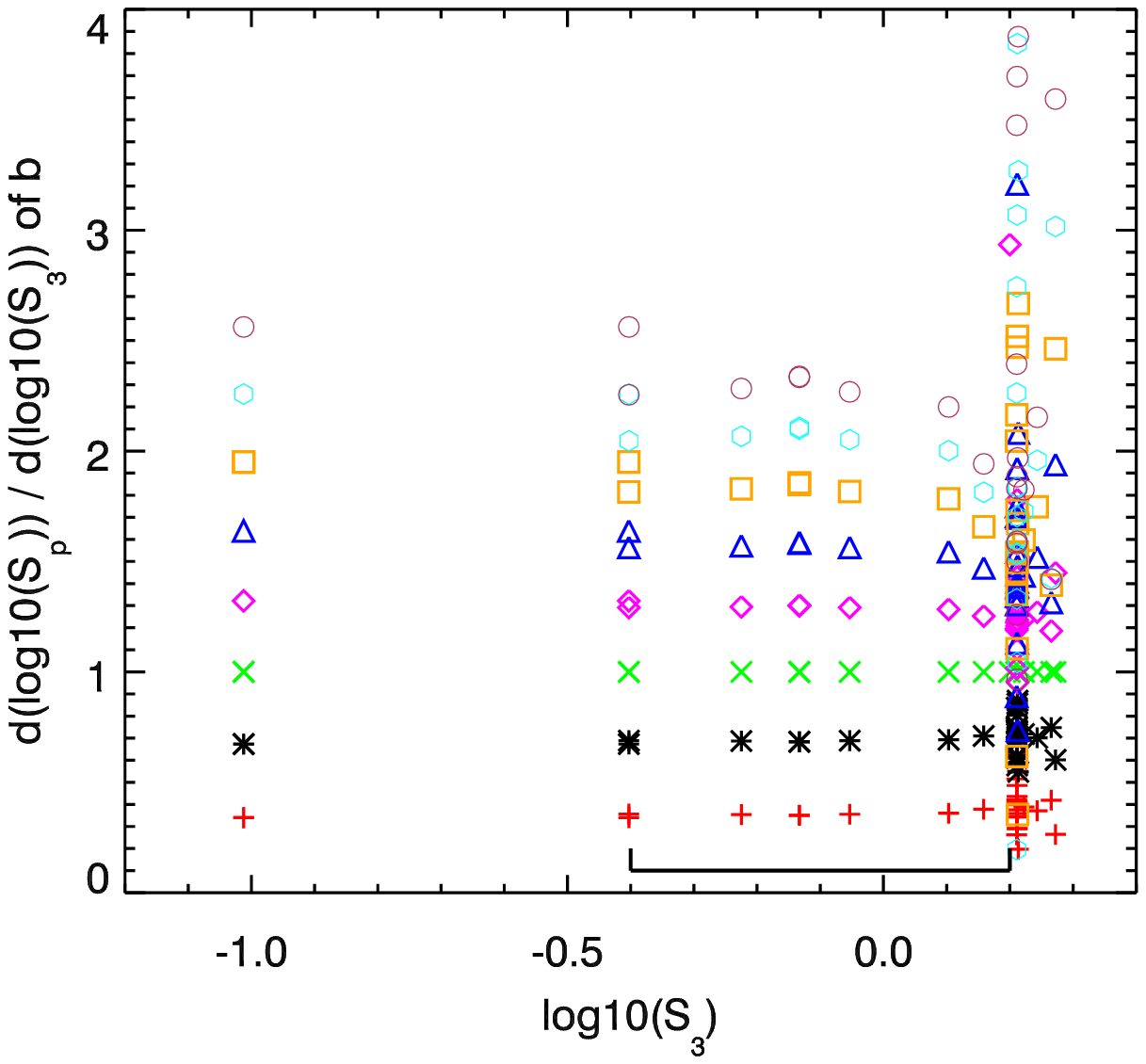}
B2)\includegraphics[width=4.5cm,height=4.5cm,viewport=43 8 410 330,clip]{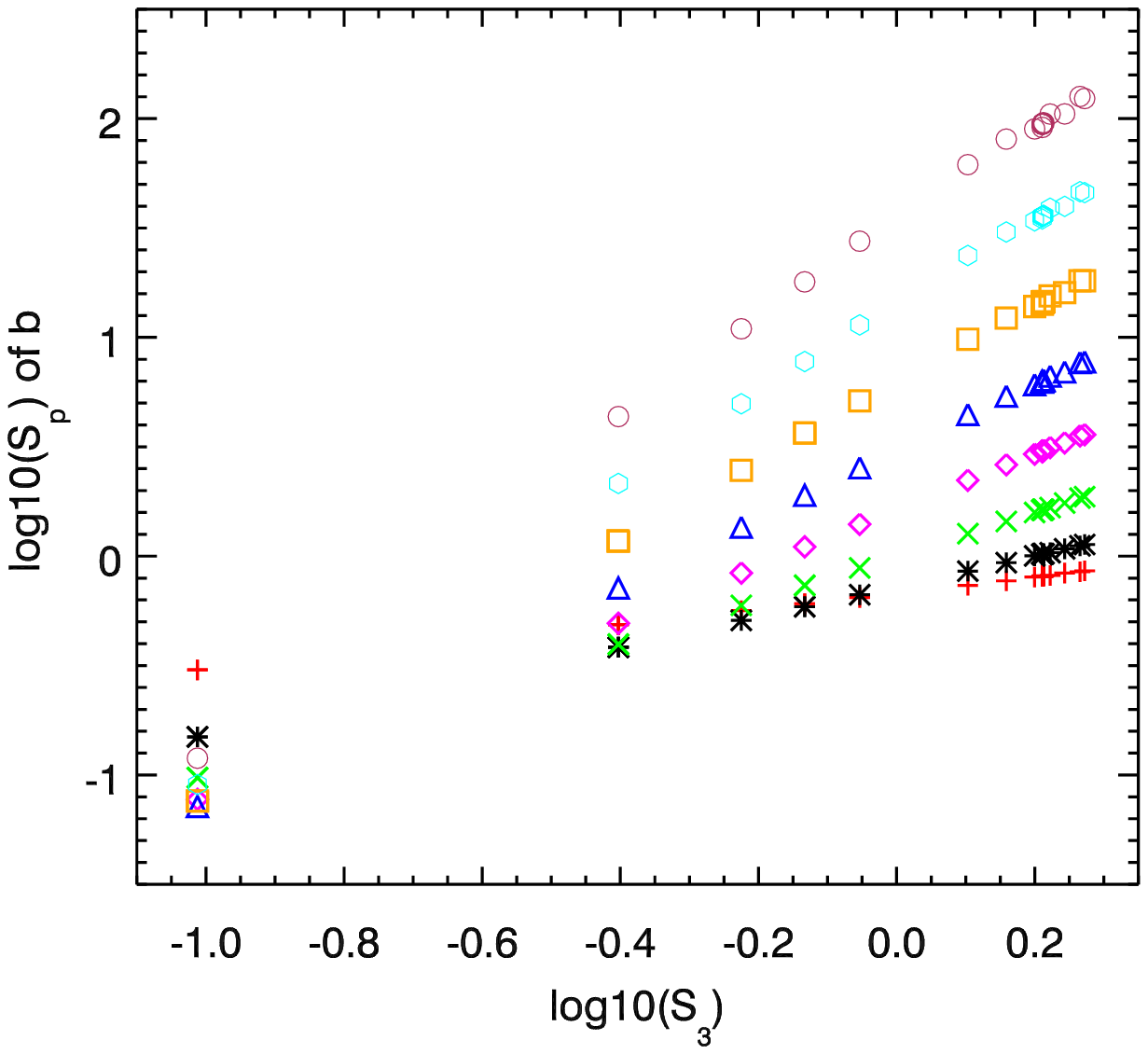}\\
B3)\includegraphics[width=4.5cm,height=4.5cm,viewport=43 8 410 330,clip]{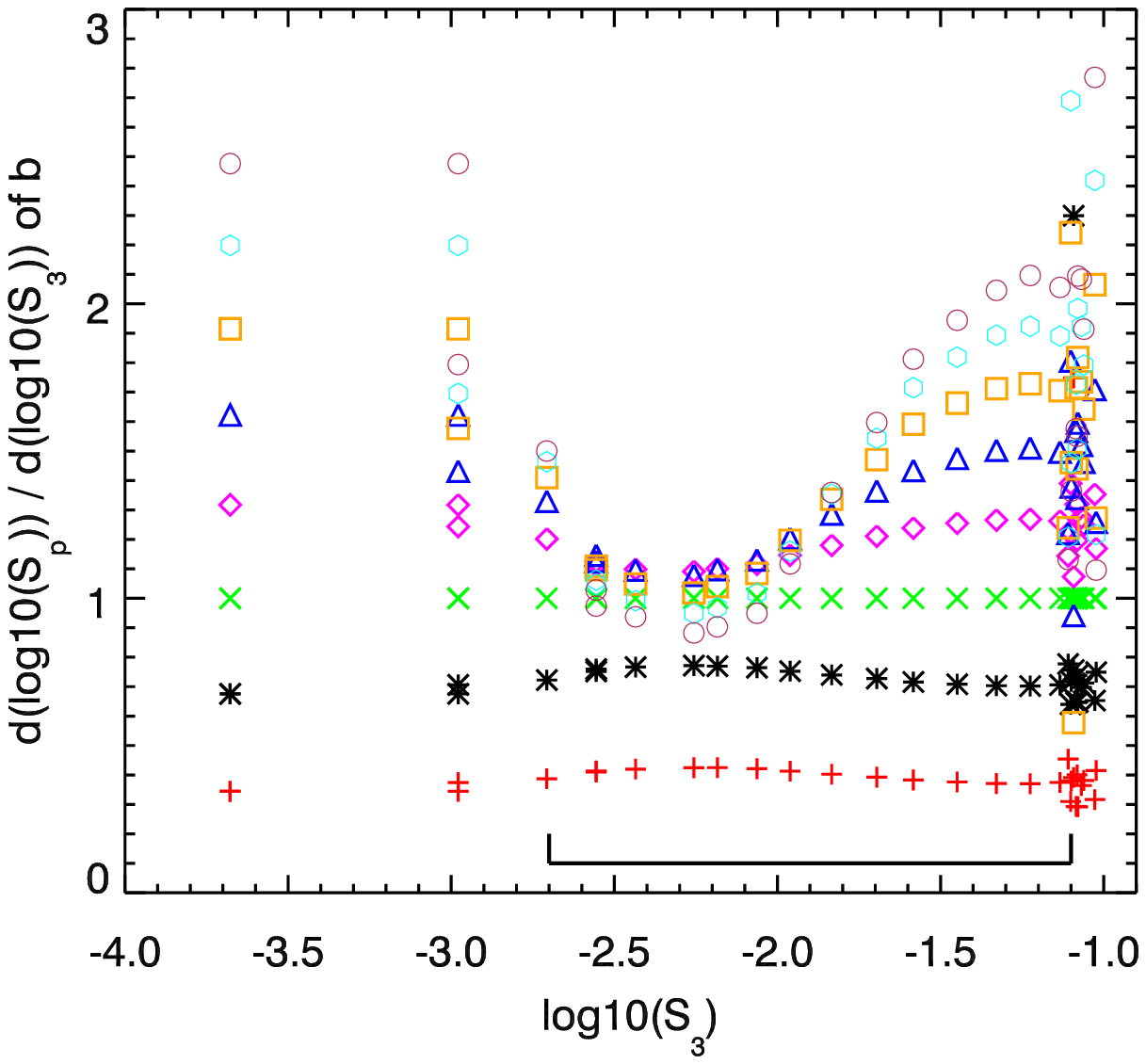}
B4)\includegraphics[width=4.5cm,height=4.5cm,viewport=43 8 410 330,clip]{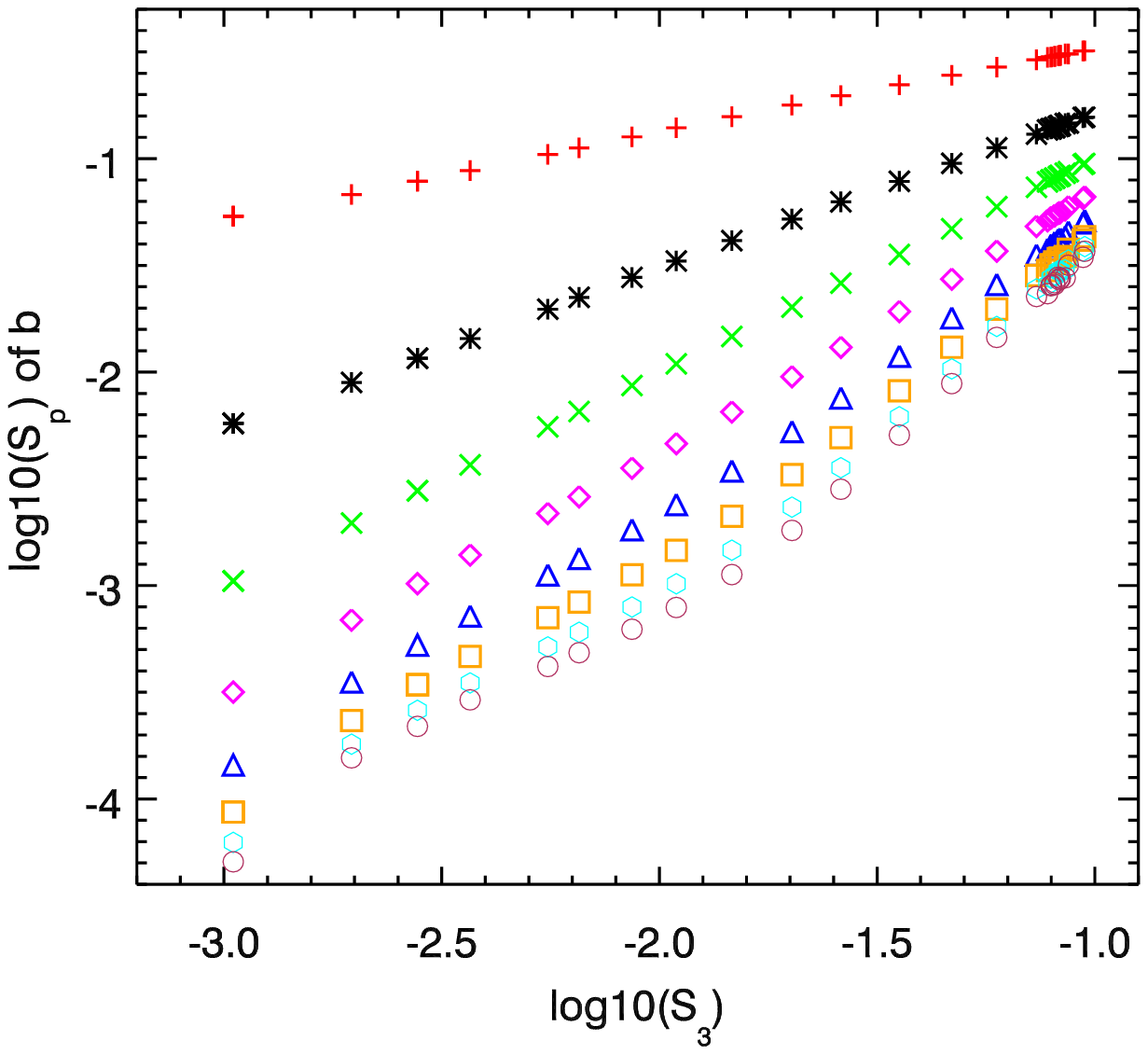}\\
B5)\includegraphics[width=4.5cm,height=4.5cm,viewport=43 8 410 330,clip]{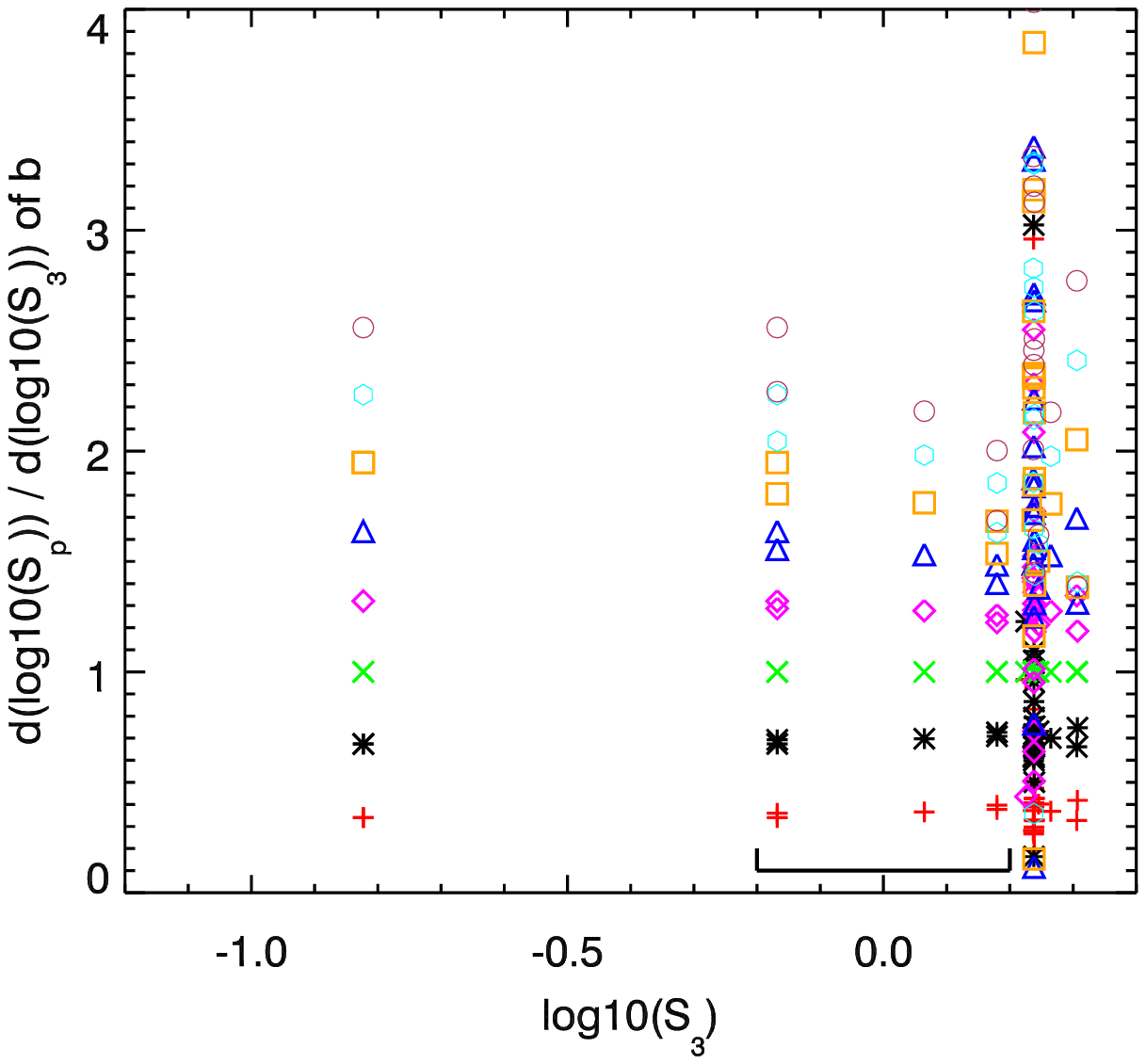}
B6)\includegraphics[width=4.5cm,height=4.5cm,viewport=43 8 410 330,clip]{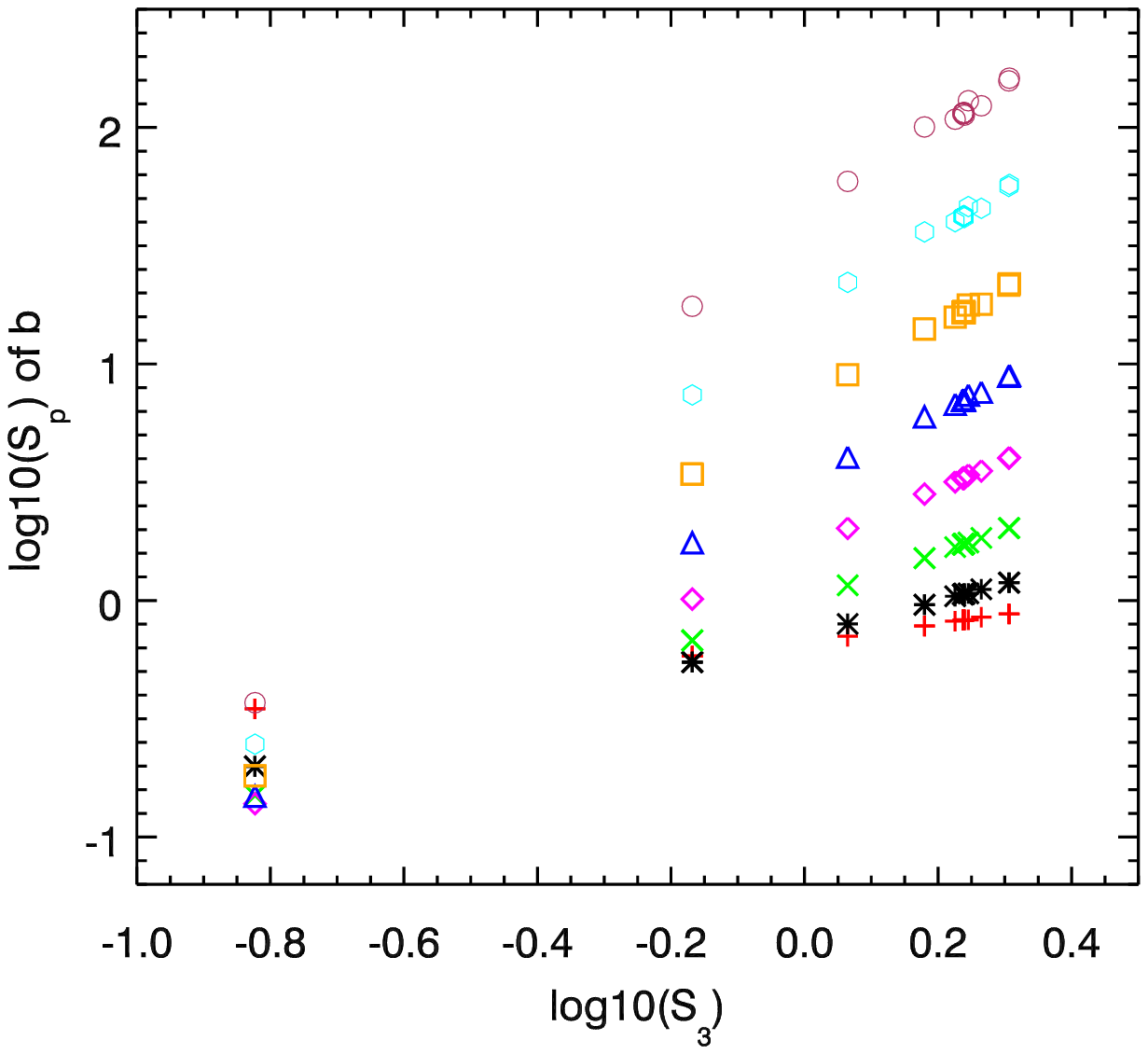}\\
B7)\includegraphics[width=4.5cm,height=4.5cm,viewport=43 8 410 330,clip]{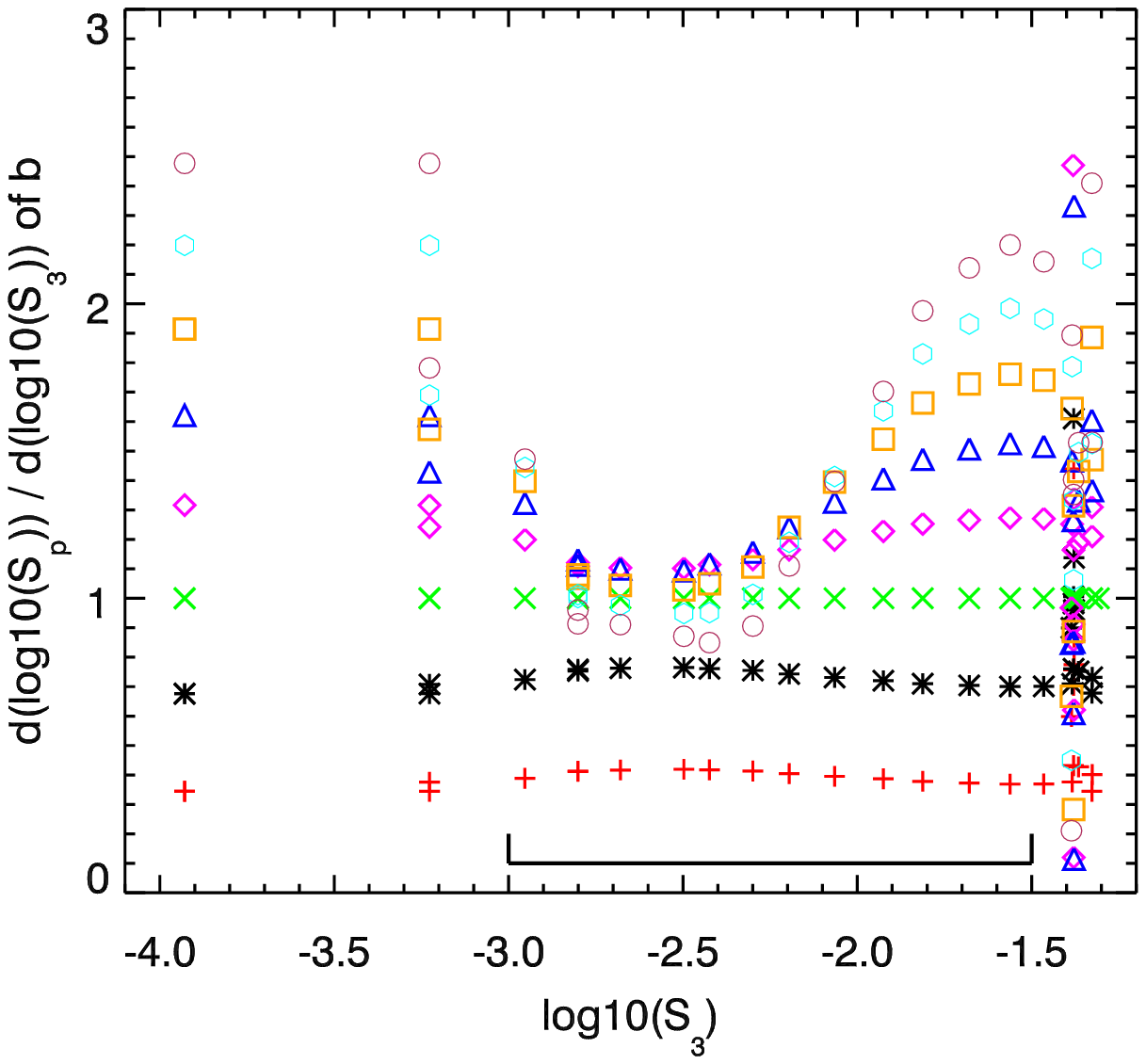}
B8)\includegraphics[width=4.5cm,height=4.5cm,viewport=43 8 410 330,clip]{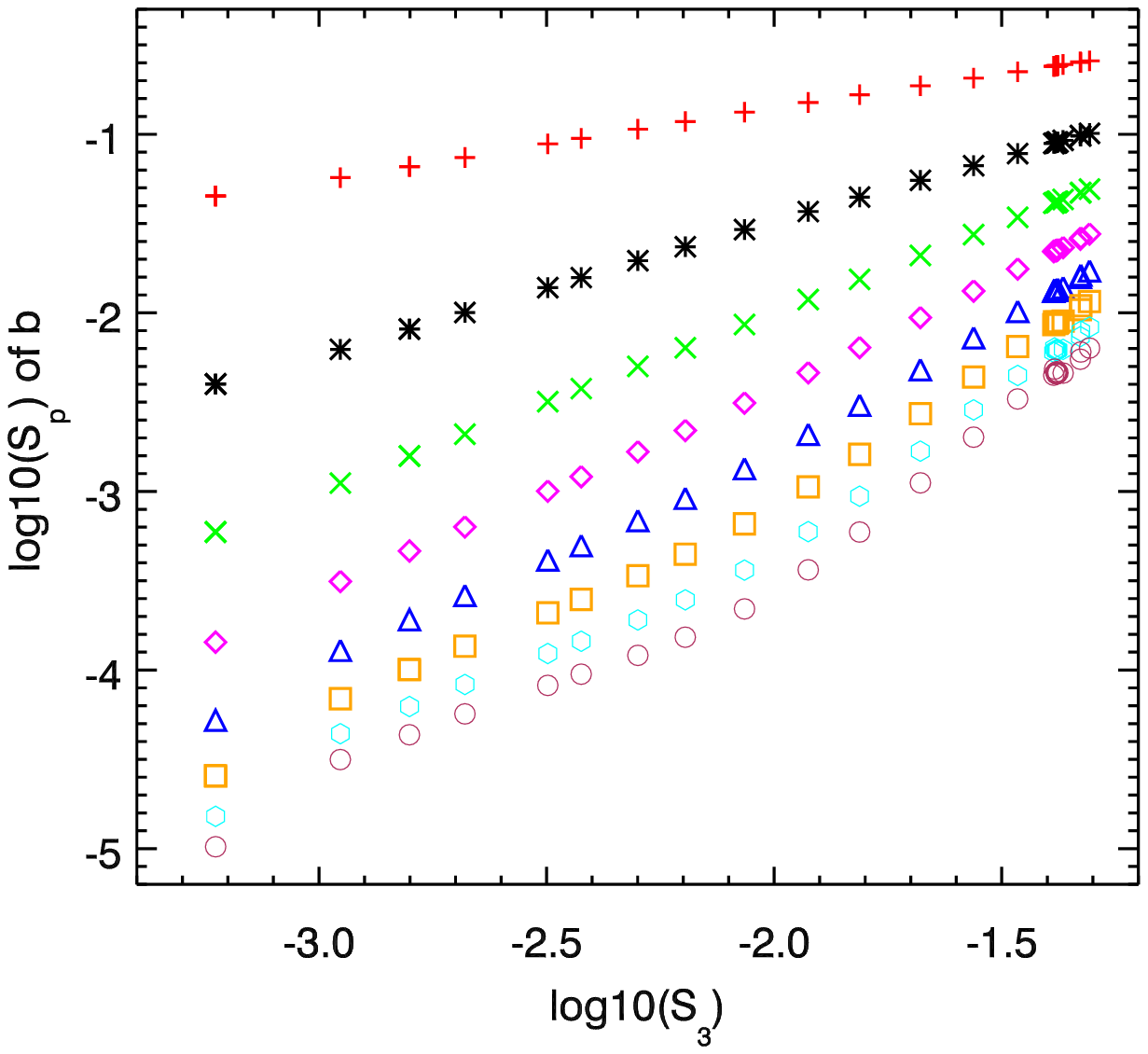}
\caption{Structure functions calculated from simulations for magnetic field for cases 3bi (B1 and B2), 3bf (B3 and B4), 3ci (B5 and B6) and 3cf (B7 and B8). B1, B3, B5 and B7 : $\frac{dlog10(S_{p})}{dlog10(S_{3})}$ Vs log10($S_3$) for Local Slope Analysis (LSA). B2, B4, B6 and B8: log10($S_p$) Vs log10($S_3$) for Extended Self Similarity Analysis (ESS). Order($\it{p}$): Color and Symbol: 1: red $+$, 2: black $*$, 3: green $\times$, 4: magenta $\diamond$, 5: blue $\vartriangle$, 6: orange $\square$, 7: cyan $\hexagon$ and 8: maroon $\circ$. [Color version of the figure is available  online]}
\end{center}
\end{figure}
\end{document}